\documentstyle[prl,aps,preprint,amssymb,noheadings]{revtex}
\begin{document}
\newcommand{\k}{\kappa}                                                       
\newcommand{\p}{\partial}                                                     
\newcommand{\Y}{{\cal Y}}
\newcommand{\R}{{\cal R}}
\newcommand{\CP}{{\cal P}}
\newcommand{\T}{{\cal T}}
\newcommand{\CQ}{{\cal Q}}
\newcommand{\CS}{{\cal S}}
\newcommand{\CH}{{\cal H}}
\newcommand{\G}{{\cal G}}
\newcommand{\TT}{\T_{11}}
\newcommand{\RR}{\R_{11}}
\newcommand{\PSbox}[3]{\mbox{\rule{0in}{#3}\includegraphics{#1}\hspace{#2}}}
\clubpenalty=10000
\widowpenalty=10000
\brokenpenalty=10000
\interdisplaylinepenalty=5000
\predisplaypenalty=10000
\postdisplaypenalty=100
\tolerance=1000

\tighten

\title{BOUND STATES AND THRESHOLD RESONANCES IN 
QUANTUM WIRES WITH CIRCULAR BENDS \thanks
{This work is supported in part by funds provided by the U.S.
Department of Energy (D.O.E.) under cooperative
agreement \#DF-FC02-94ER40818 and \#DE-FG02-92ER40702
and in
part by funds provided by the National Science Foundation under grant
\# PHY 92-18167}}

\author{K. Lin\footnote{Present address:  Department of Physics, 
366 LeConte Hall, University of California at Berkeley, Berkeley, CA 94720} }

\address{Center for Theoretical Physics, Laboratory for Nuclear Science \\
and Department of Physics \\
Massachusetts Institute of Technology, Cambridge, Massachusetts 02139 \\ 
{~}}{\centering and}

\author{R. L. Jaffe}

\address{Center for Theoretical Physics, Laboratory for Nuclear Science \\
and Department of Physics \\
Massachusetts Institute of Technology, Cambridge, Massachusetts 02139 \\
and\\
Lyman Physics Laboratory, Harvard University\\
Cambridge, Massachusetts 02138 \\
{~}}{}

\date{MIT-CTP-2504 ~~~ HUTP-95/A052 Submitted to
{\it Physical Review B} ~~~ November 1995}
\maketitle

\begin{abstract}

We study the solutions to the wave equation in
a two-dimensional tube of unit width comprised of two
straight regions connected by a region of constant curvature.  
We introduce a numerical method which permits high accuracy at high 
curvature.
We determine the bound state energies as well as the
transmission and reflection matrices, ${\cal T}$ and ${\cal R}$ and
 focus on the nature of 
the resonances which occur in the vicinity of channel thresholds.
We explore the dependence of
these solutions on the curvature of the tube and angle of the bend
and discuss several limiting cases where our numerical results confirm
analytic predictions. 

\end{abstract}
\break
\section{Introduction}

 The problem of a quantum particle
confined to a two-dimensional tube consisting of two straight regions connected
by a region of arbitrary curvature but constant width, $d$,
has aroused interest in recent years because of its
applications in both the physics of small devices and
electromagnetic waveguides.\cite{JK,TT,Review}
Related geometries such as crossed wires\cite{SRW} and elbows\cite{Lond} have 
also been studied extensively.  The primary focus has been on the existence 
of unanticipated bound states and on the ``adiabatic'' limit where the 
radius of curvature is always much greater than the width of the 
tube.\cite{Exner}  Goldstone 
and Jaffe proved that such a tube always has a bound state, with 
energy below the continuum threshold at 
$E={\hbar^2\over 2m}{\pi^2\over d^2}$.\cite{GJ}

\begin{figure}
\centering
\PSbox{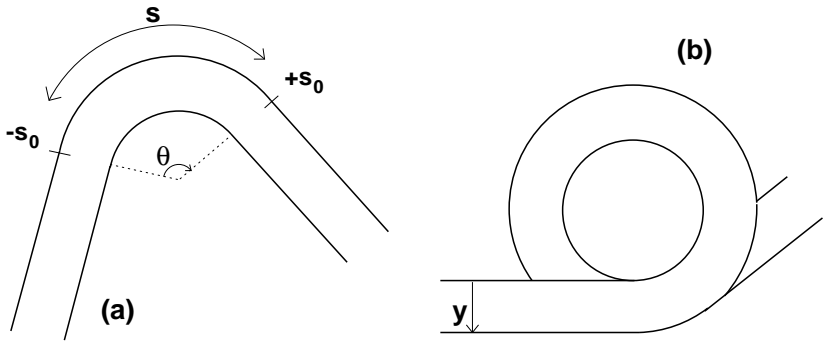 hscale=64 vscale=64 voffset=-368}{4.31in}{1.91in}
\caption{An infinite two-dimensional tube of unit width comprised of two
straight regions connected by a region of constant curvature
$\kappa$. (a) $\theta < \pi$. (b) $\theta > 2\pi$.}
\label{tube}
\end{figure}

In this paper, we study the specific
case where the curved region is an arc of a circle.  This case can be solved 
numerically with relative ease, providing a laboratory in which to explore 
phenomena we believe to be quite general but difficult to demonstrate
in arbitrary geometries.  We take the
tube to have unit width.  We use the arc length, $s$, along the outer 
boundary of the tube as one coordinate, and the distance $y$, along 
the normal from the outer boundary toward the center of the circle
as the other coordinate.
We mark the boundaries of the
curved region as $\pm s_0$.  The angle subtended by the arc of the 
curved region is defined to be $\theta$, and it satisfies the relation
$\theta = 2\k s_0$, where $\k$ is the curvature (here a constant).
Figure \ref{tube} shows two examples of such a tube.  Notice that we 
allow ourselves to consider tubes with $\theta>\pi$ although such 
configurations require excursions from strictly planar geometry.
Also, we choose units such that $\hbar^2/2m=1$ making $E=k^2$ dimensionless.  

Scattering in a bent tube is a relatively simple example of a
multichannel problem.  Incoming and outgoing waves are labelled by an
integer, $n$, the number of non-trivial nodes in the transverse
wavefunction, $\sin n\pi y$.  The scattering is described by
transmission and reflection matrices $\T_{mn}$ and $\R_{mn}$ whose
dimension grows as each successive channel threshold (at
$E_n=n^2\pi^2$) is passed.  In Ref.~\cite{GJ} it was pointed out that
the same argument which demonstrates the existence of a bound state,
also proves that there is a ``quasibound'' state just below each
channel threshold.  The quasibound state below the $N^{th}$ channel
threshold would be stable and normalizable if all lower channels
($n<N$) were artifically closed.  Such quasibound states should appear
as resonances in the open channel scattering amplitudes.

Our aim in considering this solvable special case is to explore the
systematics of the bound states as they depend on $\k$ and $\theta$,
and to elucidate the character of the channel threshold resonances.
Other workers have studied quantum wires with circular bends.  Lent
studied scattering in tubes with $\theta = \pi/2$ and showed that even
for the largest curvature tractible there is very little reflection
except at values of the energy just above the threshold.\cite{Lent}
Sols and Macucci have confirmed this result and discovered the
presence of resonances associated with quasi-bound states just below
propagation thresholds.\cite{SM} Moreover, they found that at large
angles, more than one bound state may develop, and that the binding
energy of these bound states increases as the curvature or angle is
increased.

In \S II, we establish a formalism for solving for both bound and 
scattering states in tubes with circular bends.  We begin traditionally -- by 
representing the solution in the curved region in terms of
Bessel functions and imposing the standard continuity conditions at
$s=\pm s_0$.  Normally one would match the open channels to incoming and
outgoing waves and the closed channels to {\it falling exponentials\/}.
The matching conditions generate a linear algebra problem, which 
leads eventually to an eigenvalue problem (for the bound states) and a 
solution for $\T_{mn}$ and $\R_{mn}$ (for the scattering problem). 
However, we find that this approach leads to serious convergence 
problems at high curvature.  At this point we abandon the traditional approach
and formuate a novel method (as far as we are aware) where we match the
interior wavefunction to {\it both rising and falling exponentials\/} 
in closed
channels.  We then seek solutions by minimizing the coefficients of the rising
exponentials.  We argue that this approach should converge more rapidly
and reliably.  We believe this method has wider application than the 
problem at
hand.  Also in this section we derive a 
small $\k$ approximation to the transmission amplitude in regions far from 
thresholds and
resonances.

In \S III we study the
convergence of our method of calculating both scattering and bound states.  We
show that it is more convergent and more efficient than the traditional method
(of allowing only falling exponentials in closed channels).

In \S IV, we present results for the bound state eigenenergies
and the scattering coefficients as functions of curvature and angle.  We show
that at fixed curvature, further bound 
states appear with increasing $\theta$. 
As $\theta\rightarrow\infty$, so does the number of bound states.  The binding
energy of the ground state approaches a limit determined by the root of a
simple Bessel equation as
$\theta\rightarrow\infty$.  For $\k=1$ the 
limit is the square of the first zero of
$J_0$ ($\approx (2.405)^2$).  The higher 
bound state energies approach the same
limit as the ground state, but more slowly.   We also find the angle at which
the first excited state appears, as a function of $\k$, and give a theoretical
explanation why this angle approaches $2\pi$ as
$\k$ approaches zero.  We also plot the scattering coefficients $\T$
and $\R$ as functions of the energy and verify the presence of
resonances associated with quasi-bound states just below the
propagation threshold for each channel.  Finally, we explore the
behavior of $\R$ and $\T$ in the complex plane.  We explain why the elastic
transmission amplitude $\T_{11}$ drops to zero at the energy of the quasibound
state just below the second threshold, $E=4\pi^2$, an effect which is 
echoed in other components of $\T$ at higher thresholds.
At the same time we verify the predictions of the small $\k$ 
approximation for low curvature.

\section{Theory}

In this section we summarize the basic formalism common to both the bound and
scattering state problems.  We then specialize to the bound state case and
finally, to the scattering case.

\subsection{General Formalism}

We are interested in the solutions to the wave equation,
$-\vec\nabla^2\psi=k^2\psi$ within a ``tube'' coordinatized by $s$ and $y$. 
$\psi$ vanishes on the boundaries, 
$y=0$ and $1$, and we require that $\psi$ and
and its normal derivative
be continuous at $s=\pm s_0$ where the curvature of
the tube jumps abruptly from zero to $\k$.  We treat the curved and straight
regions at the same time, taking  $\k=0$ for the straight region.
The coordinate differential is given by
$d\vec{x} = \hat sds(1-\k y) + 
\hat y dy$, and the differential area element is
$d^2x=dsdy(1-\k y)$.
Thus, the wave equation becomes
\begin{equation}
\kappa\frac{\p}{\p y}\psi - (1-\kappa y)\frac{\p^2}{\p
y^2}\psi - \frac{1}{1-\kappa y}\frac{\p^2}{\p s^2}\psi = (1-\kappa y)
k^2\psi.
\label{SE}
\end{equation}
We now make the standard change of variable, $\psi
= \frac{1}{\sqrt{1-\kappa y}}\phi$
so that eq.~(\ref{SE}) becomes 
\begin{equation}
-\frac{\p^2\phi}{\p y^2} - \left[\frac{1}{(1-\kappa y)^2}\frac{\p^2}{\p s^2} +
\frac{\kappa^2}{4(1-\kappa y)^2}\right]\phi = k^2\phi,
\label{modSE}
\end{equation}
If we define $r \equiv 1 - \k y$, and separate variables we recognize
the radial equation as Bessel's equation, with solutions of the form,
\begin{equation}
\psi_l(s,r) = \frac{1}{\sqrt{r}}S_l(s)\Y_l(r),
\end{equation}
where
\begin{equation}
\Y_l(r) =
\sqrt{r}\left[Y_{\nu_l}\left(\frac{kr}{\k}\right)
J_{\nu_l}\left(\frac{k}{\k}(1-\k)\right)
- J_{\nu_l}\left(\frac{kr}{\k}\right)
Y_{\nu_l}\left(\frac{k}{\k}(1-\k)\right)\right] 
\label{Radial}
\end{equation}
and
\begin{equation}
S_l(s) = \left\{ \begin{array}{l}
	    \cos \k\nu_l s\\
	    \sin \k\nu_l s
	    \end{array}
	\right..
\label{S_l}
\end{equation}
Here $J_\nu$ and $Y_\nu$ are Bessel and Neumann functions respectively.  The
linear combination of Bessel functions in eq.~(\ref{Radial}) 
satisfies $\Y=0$ at
$y=1\,(r=1-\k)$.  The order, $\nu_l$, 
must be chosen to satisfy $\Y=0$ at $y=0\,
(r=1)$.  $\nu_l$ may be either real or 
imaginary, as we discuss below. The label
$l=1,2,\ldots$ denotes the number of non-trivial nodes ($l-1$) in
$\Y(r)$.

To study the character of $\nu_l$, 
we substitute eqs.~(\ref{Radial})
 and (\ref{S_l}) back into the wave equation. We find that the
momentum satisfies the relation,
\begin{equation}
k^2 = \k^2 \nu_l^2 + \xi_l,
\label{xi-eq}
\end{equation}
where $\xi_l$ is a positive term that grows with $l$.  
Thus, for fixed $k^2$ only a
finite number of eigenvalues, $\nu_l^2$,
can be positive.  For any $k^2$ greater
than $\pi^2$ there is at least one positive $\nu^2$ -- a fact which is closely
related to the existence of a bound state.  As $k$ increases, more positive
$\nu^2$ values appear -- associated with the channel threshold resonances
discussed below.  Let $l\le\ell$
correspond to real $\nu$ and $l>\ell$ give imaginary $\nu$, where we define
$\beta\equiv i\nu$.  Then,
\begin{equation}
\psi_<(s,r) =
\frac{1}{\sqrt{r}}\sum_{l=1}^{\ell}a_l \Y_l(r) \left\{ \begin{array}{l}
				\cos \k\nu_l s\\
				\sin \k\nu_l s
				\end{array} \right\} +
\frac{1}{\sqrt{r}}\sum_{l=\ell+1}^{\infty} a_l \Y_l(r) \left\{ \begin{array}{l}
				\cosh \k\beta_l s\\
				\sinh \k\beta_l s
				\end{array} \right\}.
\label{psiin}
\end{equation}
We have added the subscript ``$<$'' to 
$\psi$ to denote the solution inside the
circular region.

Depending on $k^2$, the exterior solutions, 
$\vert s\vert>s_0$, either oscillate
or vary exponentially with $s$.  The general exterior solution is of the form,
\begin{equation}
\psi_> = \sum _{m=1}^{\infty} (b_{m}e^{-\gamma_{m}(s-s_{0})} +
c_{m}e^{\gamma_{m}(s-s_0)})\sin m\pi y.
\label{psiout}
\end{equation}
where
\begin{equation}
\gamma_m=\sqrt{m^2\pi^2-k^2}.
\label{gamma}
\end{equation}
When $k^2>m^2\pi^2,\,\gamma_m$ becomes imaginary and the solutions become
oscillatory.  In that case we define $\gamma_m\equiv ik_m$ with $k_m>0$.

We now apply continuity conditions at $s=s_0$, recalling that in the 
interior region the normal derivative of $\psi_<$ is
\begin{equation}
\hat{s} \cdot \vec{\nabla} \psi_< =  \frac{1}{r}\frac{\p}{\p s}\psi_<,
\end{equation}
and we find
\begin{eqnarray}
b_{m} + c_{m}&=&2 \sum_{l=1} a_{l} S_{l}(s_{0}) M_{ml}\nonumber\\
-b_{m} + c_{m}&=&\frac{2}{\gamma_{m}} \sum_{l=1} a_{l}
 \frac{dS_l}{ds}\left.\right|_{s=s_0}
N_{ml}
\label{match}
\end{eqnarray}
where $M_{ml} = \int_{0}^{1} dy \sin m\pi y \frac{\Y_{l}}{\sqrt{r}}$
and $N_{ml} = \int_{0}^{1} dy \sin m\pi y \frac{\Y_{l}}{r^{3/2}}$.
Eqs.~(\ref{match}) give an infinite dimensional matrix linear algebra
problem to solve for the energies of bound states or the
${\CS}$--matrix for scattering states.  In practice we must truncate
the sum over internal and external channels. We superpose $L$ internal
solutions and attempt to match to $M$ external channels.

When $k^2<N^2\pi^2$ the $\{c_m\}$ with $m > N$ are coefficients of
exponentially rising solutions.  Because these solutions are not normalizable,
standard practice is to dismiss them as unphysical and set $\{c_m\}$ for 
$m > N$ to zero from the beginning.  
This leads to an eigenvalue condition for
$k$ which has unique solutions only when $L = M$, 
{\it i.e.\/} one is forced  
to use channel spaces of the same dimension both inside and outside the
region of curvature.  As the dimension of this channel space
$(L=M)$ is increased, the accuracy of the solution should improve.  After
all, it corresponds to taking more terms in the approximation to the
wavefunction.  However, we 
have found that this straightforward approach fails at
high curvature ($\k > 0.6$).  As the channel space dimension is increased, the
solution does not seem to converge.  We believe that this is due to poor
matching at the boundary
$s = s_0$:  We expect that an interior 
function with $n$ nodes should match to a
superposition of exterior functions dominated by terms with $n$, $n
\pm 1$, and $n \pm 2$, {\it etc.\/} nodes.  However, the $L=M$ restriction
forbids the introduction of any exterior functions with higher frequencies
than the interior functions, so that the $n+1$, $n+2$ node terms are
not present.  Clearly it would be advantageous to allow $M > L$.  However, a
simple count of the number of degrees of freedom shows that the there is no
solution under these circumstances.

To solve this problem, we have implemented a novel approach: we retain the
rising exponentials
and seek first the set of $\{a_l\}$ and next  
the value of $k$ for which
$\sum_{m=1}^M c_m^2$ is minimized.  As we shall see, this approach
removes the $L = M$ restriction, allowing us to take as many terms in
either sum as we like.  Of course, in the limit $L, M \rightarrow \infty$,
$\{c_m\} \rightarrow 0$ in the closed channels for all physically interesting
solutions, whether bound or scattering.

\subsection {Bound state formalism}

Bound states correspond to solutions of 
Eqs.~(\ref{match}) with $k^2<\pi^2$.  
%
%
Our strategy will be to allow the $\{c_m\}$ to be non-zero
and seek solutions which minimize $\sum_{m=1}^M c_m^2$.  We anticipate that 
the parity of the ground state will be even, so we choose internal solutions 
which are superpositions of 
$\cos \k\nu s$ or $\cosh \k\beta s$.  If there are 
several bound states, this symmetric {\it ansatz\/} will find the even ones.  
A corresponding {\it ansatz\/} odd under $s\leftrightarrow -s$ finds the odd 
states.  Even and odd bound states alternate as a function of the energy.

We use the method of Lagrange multipliers to minimize $\sum c_{m}^2$
subject to the constraint that $\sum_{l=1}^L a_{l}^{2} = 1$.  There is some
arbitrariness in this prescription.  The $\{a_l\}$ must be constrained in some
way, otherwise the trivial solution $\{a_l\}=0$ will follow.  The
sensitivity of our method to the form of the constraint is discussed in the
following section.  We define
\begin{equation}
\Delta(\{a_{l}\},k) \equiv\sum_{m=1}^M  
c_{m}^{2}(\{a_{l}\}) - \lambda(\sum_{l=1}^L
a_{l}^{2} - 1)
\label{delta}
\end{equation}
and seek to minimize $\Delta$ first with 
respect to the $\{a_l\}$ and then with
respect to k.  The conditions
$\frac{\p\Delta}{\p
a_l} = 0$, lead to an algebraic problem, with a solution
$\{a_l^0(k)\}$.  Substituting back into eq.~(\ref{delta}), 
$\Delta$ reduces to a
function of $k$, whose minima we locate numerically.

First we combine eqs.~(\ref{match}) to obtain $c_m$ as a function of
the
$\{a_l\}$, 
\begin{eqnarray}
c_m&=&\sum_{l=1}^L\G_{ml}a_l,\quad {\rm where}\\
\G_{ml}&=&M_{ml}S_l(s_0) + \frac{1}{\gamma_m}N_{ml}S'_l(s_0).
\label{c(a)}
\end{eqnarray}
Next we substitute for $c_m$ in eq.~(\ref{delta}), 
and differentiate with respect to $a_l$ to obtain,
\begin{eqnarray}
\sum_{j=1}^L \Xi_{lj}a_j&=&\lambda a_l,\quad {\rm where}\nonumber\\
\Xi_{lj}&=&\sum_{m=1}^M \G_{ml}\G_{mj}
\label{Xidef}
\end{eqnarray}

This is an $L$ dimensional, matrix eigenvalue problem.  The external
channel space dimension, $M$, appears in the definition of $\Xi$, but does not
complicate eq.~(\ref{Xidef}).  Since $\Xi$ 
is a positive definite matrix, it has
real, positive eigenvalues, $\lambda^1(k)$,
$\lambda^2(k)$ {\it etc.\/}, ordered
from smallest to largest, and associated eigenvectors, $a_l^1$, $a_l^2$, 
{\it etc\/}.  Each eigenvector gives a stationary point of the functional
$\Delta$. Normalizing 
$\sum_{l=1}^L (a^i_l)^2 =1$ as required by the constraint
it is easy to see that 
\begin{equation}
\Delta(\{a_l^i(k)\},k)=\sum_{m=1}^M c_m^2(\{a_l^i\}) = \lambda^i(k),
\end{equation}
so the smallest eigenvalue gives the minimum $\Delta(k)$ at fixed
$k$.  To find the best approximation to the bound states for fixed $L$
and $M$, we search for a value of $k<\pi$ that minimizes the lowest
eigenvalue, $\lambda^1(k)$.  The magnitude of $\lambda^1(k)$ measures
the squared sum of coefficients of rising exponentials, and is a
measure of the accuracy of our solution.  There is always at least one
minimum of $\lambda^1(k)$ for $k<\pi$ --- a consequence of the general
theorem of Ref.~\cite{GJ}.  For small values of $s_0$, there is only
one minimum at $k\equiv \bar k_1$, corresponding to a ground state
energy of $\bar k_1^2$.  As $s_0$ is increased (at fixed $\k$), there
will be multiple minima of $\lambda^1(k)$ which we denote by
$\bar{k_1}$, $\bar{k_3}$, $\bar{k_5}$, {\it etc.}, with $\bar{k_p} <
\pi$.  ($\bar{k_1}$ corresponds to the ground state, $\bar{k_3}$ to
the second excited state, and so on.)  The odd-numbered excited states
($p$ even) are antisymmetric in $s$ and are obtained from an
antisymmetric {\it ansatz\/}.

We must set the number of terms in the
sums in eqs.~(\ref{Xidef}) to values such that the computations are tractable
while still retaining reasonable accuracy.  Studies of convergence and
accuracy are presented in \S III.  We emphasize, though, that the
convergence is not a trivial issue for this problem:  the default assumption,
$L=M$, fails to converge as 
$L\rightarrow\infty$ for interesting values of $\k$, forcing us to take a 
different approach.
Anticipating the results of \S III, we note that the results
presented in \S IV were obtained with $L=4$ and $M=10$,
which generated accuracy to about five decimal places.  


\subsection{Formalism for scattering states}

We now extend our formalism to handle scattering solutions.  At an energy 
$E=k^2$, the ``longitudinal'' momentum in the $n^{th}$ channel is 
$k_n=\sqrt{k^2 - n^2 \pi^2}$.
If $N^2\pi^2 < k^2 < (N+1)^2\pi^2$, then $N$ channels are ``open'':
for $n\leq N$, $k_n$ is real and wavefunctions in the external regions 
are oscillatory.  For $n>N$, $k_n$ is imaginary.  These channels are 
``closed'': we define
$ik_n = \gamma_n$; the solutions are exponential as in the
bound state problem.  Since internal and 
external channel sums cover different 
ranges we will use different indices as follows:
$i$, $j$, $k$, and $l$ label internal (curved region) solutions; and
$m$, $n$, $p$, $q$, and $r$ label external (straight region) channels.
$N$ will always denote the number of open channels.

An incident wave in the $q^{th}$ channel
will couple to all channels.  However,
the solutions in the closed channels die exponentially, so that at a
distance (s $\rightarrow \pm \infty$), we see reflection and transmission
in only the open channels.  We wish to solve for these $\R$ and $\T$
coefficients and explore their behavior as functions of the channel and
energy of the incoming wave as well as the parameters of the tube.
Unlike the bound state problem, this is not an eigenvalue problem.  There are 
many solutions (in fact $2N$) 
for each value of the energy.  Our aim is to 
find the solutions and parametrize them 
in terms of the matrices $\R$ and $\T$.

Since our internal solutions are naturally constructed as either symmetric or 
antisymmetric (in $s$), it is convenient to do likewise with the external 
solutions.  We consider the symmetric case explicitly --- from which we will 
obtain $\T+\R$.  The antisymmetric case proceeds along identical lines and 
yields $\T-\R$.  The solutions
are labelled by the indices $q$, the channel of the incoming wave, and $p$,
the channel under consideration.  The symmetric solution is the sum of the 
left- and right-incoming solutions:
\begin{equation}
s\rightarrow\pm\infty:    \sigma^q_p(s) = 
\frac{1}{\sqrt{ik_p}}(\delta_{pq}e^{-ik_p|s|} + (\T+\R)_{pq}e^{+ik_p|s|}) 
\label{TRarray}
\end{equation}
Unitarity requires $(\T+\R)^\dagger (\T+\R)=1$, and from the antisymmetric 
solution, $(\T-\R)^\dagger (\T-\R)=1$, or 
\begin{eqnarray}
\T^\dagger \T = \R^\dagger \R &  = 1,\nonumber\\
\T^\dagger \R + \R^\dagger\T & = 0.\nonumber\\
\end{eqnarray}
These constraints can be summarized by 
the requirement that the matrix
\begin{equation}
\CS = \left( \begin{array}{cc}
	\R & \T \\ \T & \R 
	\end{array}
	\right)
\end{equation}
be unitary, $\CS^\dagger \CS = 1.$  This is a special case (for Hamiltonians 
symmetric under $s\rightarrow -s$) of the general definition of the 
$\CS$--matrix,
\begin{equation}
\CS = \left( \begin{array}{cc}
	\R & \T' \\ \T & \R' 
	\end{array}
	\right)
\end{equation}
where $\R$ and $\T$ are reflection and transmission coefficients for waves 
incident from the left and $\R'$ and $\T'$ are the analogous coefficients for 
right incident waves.  Time reversal invariance requires $\CS^T=\CS$, which 
requires $\T^T=\T$ and $\R^T=\R$ for symmetric Hamiltonians.\cite{Low}

To solve for the matrices $\T$ and $\R$, we return to the basis
wavefunctions for the bound state problem, modified to accommodate the
open channels and the factor of $\sqrt{ik}$.  Note that this factor is
present only in the open channels where $n\leq N$.  Because our solution
is symmetric, we take $S_l(s)$ to be $\cos{\k\nu_l s}$,
\begin{eqnarray}
\psi^q_<(s,y) &= &\frac{1}{\sqrt{r}}\sum_{l}a_l^q Y_l(r) S_l(s) \nonumber \\
\psi^q_>(s,y)& = &\sum_{n=1}^{N}\frac{1}{\sqrt{ik_n}}(b^q_{n}e^{-ik_n(s-s_0)} 
+c^q_{n}e^{ik_n(s-s_0)}) \sin{n\pi y}  \nonumber \\ 
  & + & \sum_{n=N+1}^M(b^q_{n}e^{-\gamma_n(s-s_0)} 
+c^q_{n}e^{\gamma_n(s-s_0)})
 \sin{n \pi y}.
\label{psiarray}
\end{eqnarray}
By comparing 
Eqs. (\ref{TRarray}) and 
(\ref{psiarray}) we can identify $\T+\R$ in terms of $b$ and $c$
for open channels,
\begin{eqnarray}
b_p^q &= &\delta_{pq} e^{-ik_ps_0} \nonumber \\
c_p^q &= &(\T+\R)_{pq} e^{ik_ps_0}
\label{bcdef}
\end{eqnarray}
for $p,q \leq N$.  By analysis similar to the bound state problem,
\footnote{
It is convenient to redefine the internal 
wavefunction $S_l$ in the case where 
$\nu_l$ is imaginary.  Instead of $\cosh \k\beta_l$ we use
$\frac{1}{2}[e^{-\k\beta_l (s_0-s)} +
 e^{-\k\beta_l (s_0+s)}]$.  This amounts to multiplying 
by a constant $e^{-\k\beta_l s_0}$, which will not change our 
result but keeps the numbers tractably small when $s_0$ is large.
We must now examine {\em where} $\nu_l$ is actually imaginary.  For real $\nu$
the s-dependence of $\psi$ takes the form $\cos \k\nu s = \frac{1}{2}
(e^{i\k\nu s} + e^{-i\k\nu s})$.
By strict analogy with the straight regions, we would then expect real 
$\nu$ for open channels $l \leq N$ and imaginary $\nu$ for $l>N$.  However,
the energy thresholds for propagation are actually {\em lower} in the
curved region \cite{SM}, so that we will find, for example,
$\nu_2$ real beginning not at $k^2 = 4\pi^2$, but at $k^2 = 4\pi^2 - 
\epsilon_2$, where $\epsilon_2$ is a number 
that we can find numerically because we
expect $\nu=0$ at the threshold.} 
continuity at $s=s_0$ shows that
\begin{equation}
c^q_n = \sum_{l=1}^L \G_{nl}a_l^q
\label{c_n-def}
\end{equation}
and
\begin{equation}
b^q_n = \sum_{l=1}^L \CH_{nl}a_l^q
\label{b_n-def}
\end{equation}
where
\begin{equation}
\G_{nl} = \left\{ \begin{array}{ll}
\sqrt{ik_n} S_l(s_0) M_{nl} + \frac{1}{\sqrt{ik_n}} 
S_l^\prime (s_0) N_{nl} & n \leq N\\
S_l(s_0) M_{nl} + \frac{1}{\gamma_n} S_l^\prime(s_0) N_{nl} & n > N
	\end{array} \right.
\end{equation}
and
\begin{equation}
\CH_{nl} = \left\{ \begin{array}{ll}
\sqrt{ik_n} S_l(s_0) M_{nl} - 
\frac{1}{\sqrt{ik_n}} S_l^\prime (s_0) N_{nl} & n \leq N\\
S_l(s_0) M_{nl} - \frac{1}{\gamma_n} S_l^\prime(s_0) N_{nl} & n > N
	\end{array} \right.
\end{equation}

As in the bound state case, tradition would dictate setting the coefficients 
of growing exponentials in closed 
channels to zero {\it ab initio\/}.  We seek 
to improve convergence and accuracy by 
instead keeping the growing exponentials 
and minimizing their contributions.
We have found that the best
convergence for high $\k$ is obtained by minimizing the {\em ratio} of
the growing contributions relative to the falling contributions, {\it i.e.\/}
minimizing $\sum_{n>N}|c_n|^2$ subject to the constraint that 
$\sum_{n>N}|b_n|^2 = 1$.

Because scattering boundary conditions are complex ({\it viz.\/} 
eq.~(\ref{bcdef})), we must be more careful about the reality properties of 
the variables we consider.  Certainly $c_n$ and $b_n$ are complex in open 
channels.  Likewise, the matrices $\CH$ and $\G$ are complex, although they 
are clearly real for $n>N$.  
Our manipulations will be simplified considerably 
by using matrix notation.  Even 
though $\G$ and $\CH$ are not square matrices, 
there should be no ambiguity in the definition of quantities such as 
$\G^\dagger \CH$, {\it etc.\/}.  
We define a matrix projection onto the closed 
channels: $\CP_{mn} = 0$, except 
$\CP_{mm}=1$ when $M\ge m>N$.  Note that $\CP$ 
is an $M$ dimensional matrix in the external channel space.

For an incoming wave in the $q^{th}$ channel (suppressing the $q$
superscript), we can once again define the projection onto rising 
exponentials, and implement the constraint $\sum_{n>N}|b_n|^2 = 1$ by a 
lagrange multiplier,
\begin{equation}
\Delta({\vec a}, k)\equiv\vec a^\dagger \Xi\vec a - \lambda(\vec a^\dagger
\Omega \vec a - 1)
\label{scatterdelta}
\end{equation}
where $\Xi\equiv \G^\dagger\CP\G$ and $\Omega\equiv \CH^\dagger\CP\CH$.
Note that $\Xi$ and $\Omega$ are both real, symmetric matrices in 
the internal ($L$--dimensional) space.  In 
practical calculations, $L$ is small (compared to the 
external channel space dimension, $M$), so these matrices are relatively easy 
to manipulate.

It is convenient to work in the basis of eigenvectors of 
$\Omega$.  Let $\{\vec v^k, k=1,L\}$ be the orthonormal eigenvectors 
of $\Omega$ corresponding to (real, positive) eigenvalues $\{\omega_k\}$,
\begin{eqnarray}
\Omega \vec v^k&=&\omega_k \vec v^k\nonumber\\
\vec v^{k\dagger}\cdot\vec v^l&=&\delta_{kl}.
\end{eqnarray}
We rewrite $\vec a$ in the basis of the $\{\vec v^k\}$, using 
$\sqrt{\omega_k}$ as a metric,
\begin{equation}                                                              
\vec a = \sum_{k=1}^L \frac{1}{\sqrt{\omega_k}}\eta_k \vec v^k.
\label{a-expansion}
\end{equation}
Inserting this definition back into eq.~(\ref{scatterdelta}), we obtain
\begin{equation}
\Delta(\{\vec\eta\}, k)= \vec\eta^{\,\dagger}\,\widetilde{\Xi} \,\vec\eta - 
\lambda(\vec\eta^{\,\dagger}\,\vec\eta - 1).
\end{equation}
where
\begin{equation}
\widetilde\Xi^{kl}= {1\over \sqrt{\omega_k\omega_l}} \vec v^{k\dagger}\,
\Xi\,\vec v^l
\end{equation}
Now, the condition $\frac{\p\Delta}{\p \vec a^\dagger}=0$ becomes 
\begin{equation}
\frac{\p\Delta}{\p\vec\eta^{\,\dagger}} = 0 \Longrightarrow 
\widetilde{\Xi}\,\vec\eta = \lambda\,\vec\eta,
\end{equation}
an $L$ dimensional eigenvalue problem with solutions $\{\vec\eta^q\,{\rm 
and}\,\lambda^q,\,{\rm for}\,q=1,L\}$.  As in the bound state case, the 
eigenvalue $\lambda^q$ is the value of $\Delta$ at the stationary point.  
Small $\lambda^q$ therefore correspond to solutions to the wave
equation with negligible projection on exponentially growing solutions in the 
closed channels and are therefore physically interesting.

The number of physically significant solutions to this eigenvalue problem 
depends on the energy, $k^2$.  Below the lowest threshold, $k^2<\pi^2$, we do 
not expect (and did not find)  any solutions with small 
$\Delta$ except at the energies of bound states.  When one channel is open 
($\pi^2 \leq k^2<4\pi^2$), 
there should be a single symmetric solution to the wave 
equation for each energy.  So we expect to find one eigenvalue, $\lambda^1$, 
much smaller than the rest. When a second channel opens, there are now two 
symmetric solutions, so we expect to find two small eigenvalues, $\lambda^1$
and $\lambda^2$. We conclude that we must expand the vector $\vec a$ as a 
linear combination of the $N$ eigenvectors associated with the lowest $N$ 
eigenvalues when $N$ channels are open. 
We now expand the vector $\vec a$ in terms of the lowest $N$ eigenvectors 
$\vec \eta^r$ of $\widetilde{\Xi}$:
\begin{equation}
\vec a= \sum_{r=1}^N d_r \sum_{k=1}^L \frac{1}{\sqrt{\omega_k}}\eta_k^r\vec 
v^k,
\end{equation}
where $d_r$ is the expansion coefficient and the sum over $k$ corresponds
to the expansion in terms of $\eta^k$ as defined in eq.~(\ref{a-expansion}).
Finally, we define the $L\times N$ matrix
\begin{equation}
{\cal M}_{jr} =  \sum_{k=1}^L \frac{1}{\sqrt{\omega_k}}\eta_k^r v^k_j
\end{equation}
so that we may write compactly
\begin{equation}
\vec a =  {\cal M}\vec d.
\label{final-a}
\end{equation}

We are now prepared to return to our identifications of $b_n^q$ and $c_n^q$
from eq.~(\ref{bcdef}).  Inserting our above expression for $a_i$ 
into eq.~(\ref{b_n-def}), we find that
\begin{equation}
d_n^q = \CQ_{nq}^{-1}e^{ik_qs_0}
\label{d-exp}
\end{equation}
where $\CQ$ is an $N\times N$ matrix defined by
\begin{equation}
\CQ = \CH{\cal M}.
\end{equation}
Substituting eqs.~(\ref{final-a}) and (\ref{d-exp}) back into
eq.~(\ref{c_n-def}) and recalling our identification with $(\T+\R)$ in
eq.~(\ref{bcdef}), we obtain at last
\begin{equation}
(\T+\R)_{pq} = \{G{\cal M}Q^{-1}\}_{pq}e^{i(k_p+k_q)s_0}.
\label{T+R}
\end{equation}
Note that in the end the matrix manipulations which are calculationally 
intensive --- finding eigenvalues, eigenvectors and inverses --- are only 
carried out on matrices of dimension $L$.  The external channel index, $M$, 
can therefore be taken very much greater than $L$ without increasing 
the computation time significantly.

The antisymmetric scattering states can be analyzed in the same fashion, 
beginning with the parameterization of the scattering states,
\begin{equation}
s\rightarrow\pm\infty:    \alpha^q_p(s)  =  
\frac{1}{\sqrt{ik_p}}(\delta_{pq}e^{-ik_p|s|} + 
(\T-\R)_{pq}e^{+ik_p|s|}) 
\label{asymin}
\end{equation}
obtained by subtracting the left- and right-incoming wave solutions.
This allows us to solve for the quantity $(\T-\R)_{pq}$, which we then
combine with eq.~(\ref{T+R}) to yield the desired matrices $\T$ and $\R$.


\subsection {
Small curvature approximation for the transmission amplitude}

When the curvature, $\k$, is small, the solutions to our problem simplify 
somewhat.  There is a quite well developed theory of the adiabatic 
approximation for a quantum wire with small and slowly varying curvature ---
$\k\ll 1$ and ${d\k\over ds}\ll 1$ --- but it does not apply to a circular 
region joining two straight regions because $\k$ jumps abruptly to zero at 
$s_0$.  Also, most of the results are restricted to
$sqrt(k^2-\pi^2)\approx 0$.  Nevertheless we can use the fact that $\k$
is small to simplify the  calculation of the {\it elastic\/} scattering
amplitudes $\T_{mm}$ and 
$\R_{mm}$ in regions away from resonances and bound states.  The result is
\begin{eqnarray}
\T_{mm} &=& \left[1+{\cal O}(\k^2)\right]
\exp\left(-ik_m\theta\over 2 +{\cal O}(k_m\k\theta)\right)\nonumber\\
\R_{mm} &=& {i\k\over 2}\sin(2k_ms_0-k_m\theta/2)\left[1+{\cal O}(\k)\right]
\exp\left(-ik_m\theta\over 
2+{\cal O}(k_m\k\theta)\right),
\label{smallkappa}
\end{eqnarray}
where $k_m=\sqrt{k^2-\pi^2}$ and these expressions are only valid when 
$k_m^2\gg\k^2$.

The derivation of these results begins from 
eq.~(\ref{modSE}), which we expand 
in a Fourier series, $\phi(s,y)=\sum_nf_n(s)\sin n\pi y$, 
\begin{equation}
D_{mn}(f_n''+{\k^2\over 4}f_n)+k_m^2f_m=0,
\end{equation}
where
\begin{equation}
D_{mn}=2\int_0^1 dy {\sin m\pi y\sin n\pi y \over (1-\k y)^2}.
\end{equation}
Since $D_{mn}$ is of order $\k$ when $m\ne n$, it is easy to see that the 
effect of the $n^{th}$ channel on a wave incident in the $m^{th}$ channel 
begins at order $\k^2$.  This confirms 
our decision {\it to work only to order 
$\k$\/}.  To this order $D_{mm}=1+\k$, which makes this system unusual, since 
the perturbation appears in lowest order 
only in the kinetic term.  In the end 
this will lead to the result that an arbitrarily small curvature can result 
in a large scattering phase provided the product $\k s_0\propto\theta$ 
is large.

Working to order $\k$ we can ignore the ``potential'' $\k^2/4$, with the 
result that
$f_m\propto\exp\pm (iq_m s)$, where 
$q_m^2=k_m^2(1-\k + {\cal O}(\k^2))$.  We now construct symmetric and 
antisymmetric combinations of these interior solutions, and match them and 
their normal derivatives to the appropriate external scattering states
eqs.~(\ref{TRarray}) and (\ref{asymin}).  Since 
channel coupling is of higher 
order in $\k$, $\T_{mm}\pm\R_{mm}$ are both 
unitary, so we parametrize them by 
phases, $\T_{mm}\pm\R_{mm}=\exp(2i\delta^m_\pm)$.  The result of the
matching  calculation is
\begin{equation}
\tan(k_ms_0+\delta^m_\pm)=(1\pm{\k\over 2}+{\cal O}(\k^2))\tan q_ms_0.
\end{equation}
Note that $q_ms_0$ and $k_ms_0$ differ by $-k_m\k s_0/2$ which need not be 
small even when $\k\rightarrow 0$ because $2\k s_0=\theta$ can be held fixed 
in this limit.  So the leading effect on $\T_{mm}$ and $\R_{mm}$ arises from 
this accumulating phase along the bending section.  The prefactor 
$(1\pm{\k\over 2})$ determines the modulus of $\T_{mm}$ and $R_{mm}$.
When we solve for $|\T_{mm}|$,
we find that effects of order $\k$ cancel, which they must do since 
$|\R_{mm}|\sim\k$ and $|\T_{mm}|^2 + |\R_{mm}|^2$ must give unity to order 
$\k^2$.  This completes our derivation.  The results, 
eqs.~(\ref{smallkappa}),
give a useful approximation to the elastic amplitudes 
over much of their range 
when $\kappa\ll 1$. 


\section{Computation and Convergence}

The straightforward approach to solving 
the wave equation by matching interior 
to exterior solutions failed for physically interesting values of curvature 
and angle, forcing us to develop a different method of calculation.  We are 
not aware of any studies of this method in the literature, and therefore 
present some details of our studies of convergence here.  Readers not 
interested in calculational methods should skip directly to the results 
presented in \S IV.

\subsection{Convergence of calculations for the bound state energies}

\begin{figure}
\centering
\PSbox{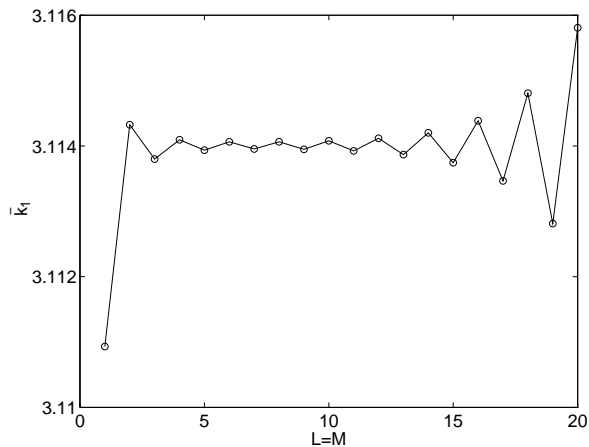 hoffset=-8 voffset=-75 hscale=42 vscale=42}{3.1in}{2.6in}
\caption{The eigenvalue $\bar{k_1}$ plotted as a function of the 
channel space in the case $L=M$, for $\k=0.8$,$\theta=\pi$.  As the
channel space is increased, the solution starts to converge but then
begins to oscillate wildly.}
\label{badconv}
\end{figure}

The traditional method of matching $L$ independent internal
solutions to $L$ independent external solutions does not converge as
$L$ increases at large values of $\k$.  This is illustrated
graphically in Figure \ref{badconv}, where we
plot the calculated ground state energy $\bar{k_1}$ as a function of
$L$ for $\kappa=0.8$ and $\theta=\pi$.  We have attributed this
failure to converge to the absence of small wavelength (in $y$) states
on the outside, necessary to accurately expand the internal states.
In order to demonstrate that this is indeed the case, we have studied
the convergence as a function of $L$ and $M$.  Table \ref{conv} shows
the effect on the calculated ground state energy $\bar{k_1}$ of
varying the channel spaces $L$ and $M$ in the case where $\k=0.8$ and
$\theta = \pi$.  The best accuracy for the least amount of computation
time is obtained by setting $M > L$.  We can see from the lower left
corner of the table that setting $M\leq L$ does not produce favorable
results, because as higher-frequency interior functions enter the
picture, we require even more high-frequency exterior functions for a
good match at the boundary $s=s_0$.

\begin{table}
\centering
\begin{tabular}{ l | llll}
$L$& $M$=5&	$M$=10&	$M$=30&	$M$=50\\\hline
1&  2.9837e-03&   2.9834e-03&   2.9834e-03&   2.9834e-03\\
2&  5.5981e-05&   5.4785e-05&   5.4635e-05&   5.4635e-05\\
3&  8.3369e-05&   7.2461e-05&   7.2181e-05&   7.2180e-05\\
4&  6.0521e-06&   9.3112e-06&   9.1729e-06&   9.1717e-06\\
5&  7.4863e-05&   1.0638e-05&   9.9300e-06&   9.9287e-06\\
6&  2.6707e-03&   1.6007e-06&   2.2744e-06&   2.2728e-06\\
7&  7.6036e-03&   4.1820e-06&   2.3406e-06&   2.3390e-06\\
8&  8.3919e-04&   3.6752e-06&   5.7210e-07&   5.7000e-07\\
9&  4.7389e-03&   9.8571e-06&   5.7590e-07&   5.7370e-07\\
10&  9.7321e-04&   6.9994e-05&   2.9000e-09&   0.0000e+00\\
\end{tabular}
\caption{Tabulation of $|\bar{k_1}(L,M) - \bar{k_1}(10,50)|$ for
$\k=0.8, \theta=\pi.$}
\label{conv}
\end{table}

We notice a few general trends in the convergence of $\bar{k_1}$.
As expected, increasing $M$ at fixed 
$L$ improves the accuracy of the result.  
Note that the larger the value of $L$, 
the larger $M$ must be taken to capture 
this improvement.  Note also that once the improvement has occured, there is 
little to be gained by further increasing $M$.   So, for example, $L=5$ 
improves dramatically as $M$ increases from $5$ to $10$, and improves little 
thereafter.  In contrast $L=9$ improves through $M=30$, and little thereafter.
Also, for fixed $M$, the convergence
worsens as $L$ is increased past the value of $M$, as we can see clearly
in the $M=5$ case.  Increasing the number of interior basis functions
should improve the accuracy of the representation of the interior
wavefunction, but apparently this is useless when the corresponding
higher-frequency terms needed to match to the outside wavefunction are
omitted, as is the case when $M \leq L$.  As can be seen from the
table, our parameters of $L=4$ and $M=10$ yield accuracy to about one
part in $10^6$.

We also note that as the number of interior basis functions $L$ is
increased, the value of $\Delta(\bar{k_1}) = \sum c_m^2$ decreases,
again provided that $M > L$.  Table \ref{conv2} illustrates this
convergence for the same parameters $\k = 0.8$ and $\theta = \pi.$ This
is to be expected; as the bound state approximation is improved, the
contributions of the exponentially rising terms should approach zero.
Although the value of $\Delta(\bar{k_1})$ does not decrease for fixed L
and increasing M, we suspect that this is a result of the increased
number of terms in the sum $\sum_m^M c_m^2$, which offsets the expected
decrease in rising exponential contributions.

For the case $L = M$, $\Delta(\bar{k_1})$ deviates from the trend in
that it is many orders of magnitude smaller than for the other cases.
This can be explained by recalling the 
traditional approach discussed in
\S II.A, where the coefficients of the rising exponential exterior
functions are set to zero from the beginning.  The resulting set of
equations has solutions only when $L = M$.  With our approach, we thus
expect to find a solution where the sum of these coefficients
$\Delta(\bar{k})$ is zero (limited only by machine precision) when $L =
M$.

\begin{table}[htbp]
\centering
\begin{tabular}{l | llll }
$L$&   $M$=5&	$M$=10&	$M$=30&	$M$=50\\\hline
1&	2.239313e-03& 2.248376e-03& 2.248825e-03& 2.248828e-03\\
2&	1.199955e-04& 1.202554e-04& 1.203799e-04& 1.203814e-04\\
3&	3.572416e-05& 4.230296e-05& 4.261721e-05& 4.261908e-05\\
4&	7.442269e-06& 8.907401e-06& 8.950761e-06& 8.952150e-06\\
5&	2.754173e-19& 4.409149e-06& 4.686405e-06& 4.688083e-06\\
6&	2.209942e-17& 1.565973e-06& 1.632015e-06& 1.633293e-06\\
7&	1.440247e-09& 7.985669e-07& 1.019960e-06& 1.021530e-06\\
8&	1.770510e-05& 3.232110e-07& 4.602362e-07& 4.613563e-07\\
9&	1.573734e-04& 1.585803e-07& 3.173988e-07& 3.188958e-07\\
10&	7.881601e-02& 2.996393e-19& 1.676304e-07& 1.685153e-07\\
\end{tabular}
\caption{$\Delta(\bar{k_1})$ for $\k=0.8, \theta=\pi$ with varying
channel space $L$ and $M$.}
\label{conv2}
\end{table}


\subsection{Convergence of different methods for scattering states}

In our study of scattering we explored further the choice of constraint
in the Lagrange multiplier problem.  One 
obvious choice is simply to constrain 
our solution to match to the asymptotic conditions in a channel of interest,
namely that the equations for $b_n^q$ 
(eq.~(\ref{b_n-def})) must be satisfied 
in the open channels.  Our equation for
$\Delta$ would then require a Lagrange multiplier for each separate channel,
leading to 

\begin{equation}
\Delta = \sum_{m=N+1}^{M}(G_{ml}^\ast a_l^{\ast})(G_{mj}a_j) -
\sum_{n=1}^{N}\lambda_n(H_{nl}a_l - b_n).
\label{method1}
\end{equation}
We note that the incoming channel label, $q$, introduced in \S II.C, 
which should appear on $\Delta, a_j,$ and $b_n$, has been suppressed for 
clarity.

Extremizing this expression for $\Delta$ yields a set of equations
from which we can determine $a_l$ and $\lambda_n$ in a straightforward
manner.  However, we have found that this method has poor convergence
for large $\k$.  Next we considered the same condition as used for the
bound states, namely eq.~(\ref{delta}) with the extension that the
coefficients $a_l$ may now be complex.  While this second method
converges for significantly higher values of $\k$ than the previous
method, we get even slightly better convergence with the procedure
outlined in \S II.C, in which we use the constraint $\sum
\vert b_n^q\vert^2 = 1$.

Table \ref{scatterconv} illustrates the convergence of $|\R_{11}|$
for different $L$ and $M$ in the case $k=6.28$, $\k=0.9$, and
$\theta=\pi/2$ for the constraints $H_{nl}a_l = b_n$ and $\sum |b_n|^2
= 1$.  This is a region --- near the second threshold at $2\pi$ and at
large $\k$ --- where convergence is problematic.  By examining
$|\R_{11}(L,M) - \R_{11}(14,24)|$ for the two choices of constraint, we see
that convergence is better for the second method.  This is primarily
due to the fact that for a given value of $L$, increasing $M$
continues to improve accuracy for a longer range in the second
method.  For example, in the case $L=10$ ,$M=12$, the two methods are
comparable, but when we reach $L=10$, $M=22$, the quantity
$|\R_{11}(L,M) - \R(14,24)|$ is a full order of magnitude smaller for
the second method.


\begin{table}
\centering

\begin{tabular}{r|rrrrrrr}
$M$ & $L=2$ & 4 & 6 & 8 & 10 & 12 & 14 \\\hline
2& 2.4008e-01 &&&&&& \\
4& 3.1798e-01 & 3.7502e-01 &&&&& \\
6& 3.2236e-01 & 1.4818e-01 & 1.2767e-01 &&&& \\
8& 3.2343e-01 & 1.6741e-01 & 4.7922e-02 & 2.4390e-01 &&& \\
10& 3.2381e-01 & 1.7536e-01 & 4.8338e-02 & 3.2088e-02 & 1.3033e-01 && \\
12& 3.2397e-01 & 1.7893e-01 & 5.2158e-02 & 1.7015e-02 & 4.8322e-02 & 3.8592e-01 & \\
14& 3.2405e-01 & 1.8067e-01 & 5.4878e-02 & 1.5969e-02 & 1.1385e-02 & 1.0571e-01 & 1.8553e-01 \\
16& 3.2410e-01 & 1.8159e-01 & 5.6600e-02 & 1.6559e-02 & 6.2273e-03 & 1.7310e-02 & 2.2753e-01 \\
18& 3.2412e-01 & 1.8209e-01 & 5.7671e-02 & 1.7235e-02 & 5.4160e-03 & 4.4498e-03 & 4.1930e-02 \\
20& 3.2414e-01 & 1.8240e-01 & 5.8343e-02 & 1.7773e-02 & 5.4195e-03 & 1.9425e-03 & 7.7122e-03 \\
22& 3.2415e-01 & 1.8258e-01 & 5.8771e-02 & 1.8168e-02 & 5.5804e-03 & 1.3692e-03 & 1.5033e-03 \\
24& 3.2415e-01 & 1.8270e-01 & 5.9049e-02 & 1.8451e-02 & 5.7499e-03 & 1.2636e-03 & 0.0000e+00 \\
\end{tabular}

(a) $|\R_{11}(L,M) - \R(14,24)|$ using the constraint $H_{nl}a_l^q =
b_n^q$ for all $n\leq N$.

\begin{tabular}{r|rrrrrrr}
$M$ & $L=2$ & 4 & 6 & 8 & 10 & 12 & 14 \\\hline
2 & 2.4129e-01 &   &   &   &   &   &   \\
4 & 2.3975e-01 & 1.2888e-01 &   &   &   &   &   \\
6 & 2.3969e-01 & 4.3621e-02 & 1.3154e-01 &   &   &   &   \\
8 & 2.3968e-01 & 3.1081e-02 & 2.3833e-02 & 1.8675e-01 &   &   &   \\
10 & 2.3968e-01 & 2.7993e-02 & 9.4324e-03 & 2.7598e-02 & 2.7725e-01 &   &   \\
12 & 2.3968e-01 & 2.7046e-02 & 5.8291e-03 & 7.0832e-03 & 4.6534e-02 & 3.6196e-01 &   \\
14 & 2.3968e-01 & 2.6712e-02 & 4.6140e-03 & 2.5538e-03 & 9.8018e-03 & 9.1442e-02 & 4.0612e-01 \\
16 & 2.3968e-01 & 2.6580e-02 & 4.1339e-03 & 1.1407e-03 & 2.4954e-03 & 1.8614e-02 & 1.7307e-01 \\
18 & 2.3968e-01 & 2.6523e-02 & 3.9258e-03 & 5.9470e-04 & 4.5970e-04 & 4.3728e-03 & 4.0580e-02 \\
20 & 2.3968e-01 & 2.6496e-02 & 3.8295e-03 & 3.5300e-04 & 2.5870e-04 & 7.7920e-04 & 9.5638e-03 \\
22 & 2.3968e-01 & 2.6482e-02 & 3.7825e-03 & 2.3630e-04 & 5.5810e-04 & 3.6960e-04 & 2.1651e-03 \\
24 & 2.3968e-01 & 2.6475e-02 & 3.7584e-03 & 1.7670e-04 & 6.9850e-04 & 8.0860e-04 & 0.0000e+00 \\
\end{tabular}

(b)  $|\R_{11}(L,M) - \R(14,24)|$ using the constraint 
$\sum \vert b_n^q\vert^2  = 1$, our actual method.
\vspace{0.25in}
\caption{Tabulation of $|\R_{11}(L,M)| - |\R_{11}(14,24)|$ 
for $k=6.28$, $\k=0.9$,
and $\theta = \pi/2$ for two different choices of constraint.}
\label{scatterconv}
\end{table}

\begin{figure}
\centering
\parbox{3in}{\PSbox{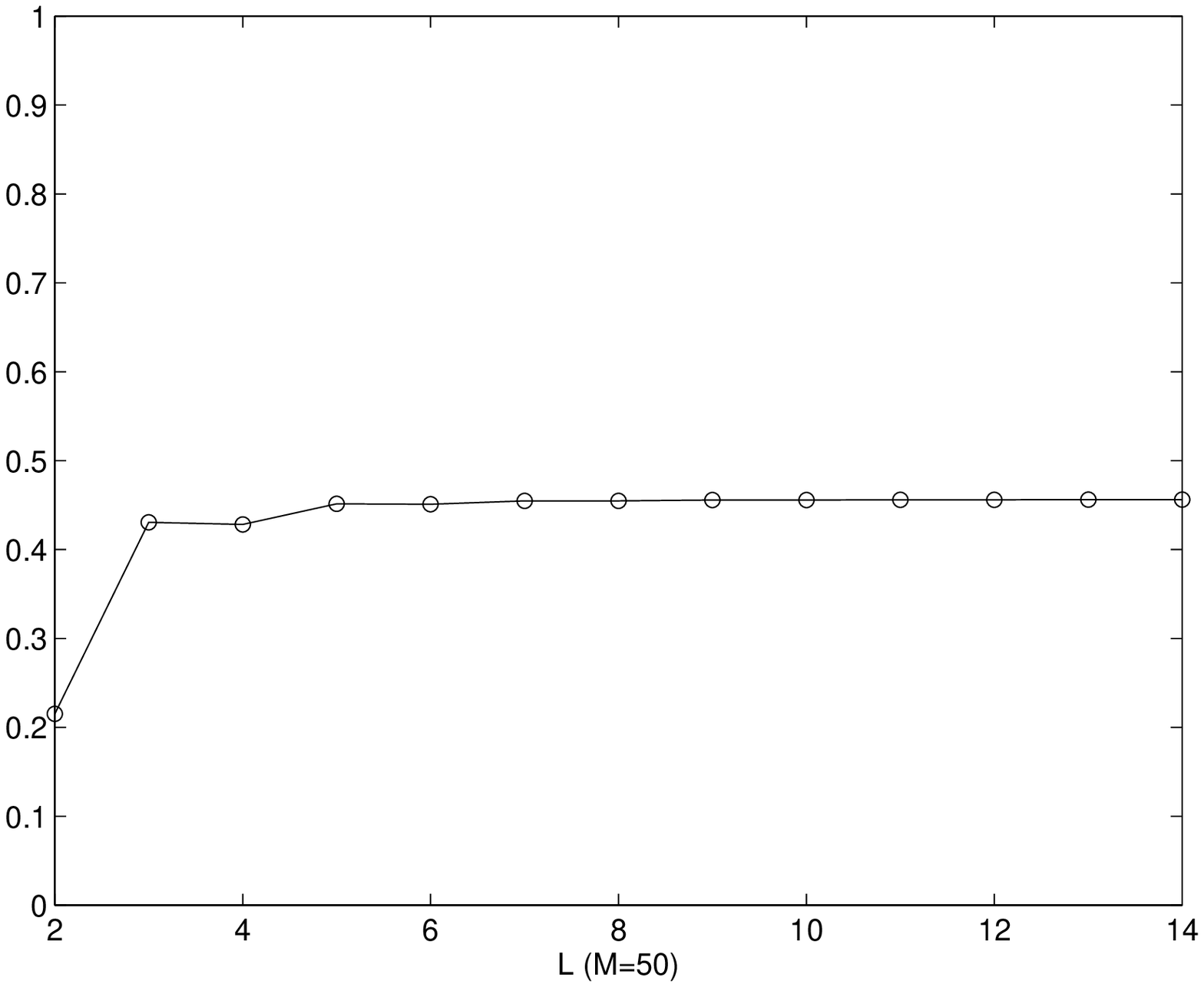 hoffset=-32 voffset=-84 hscale=45 
vscale=45}{2.97in}{2.4in}}
\parbox{3in}{\PSbox{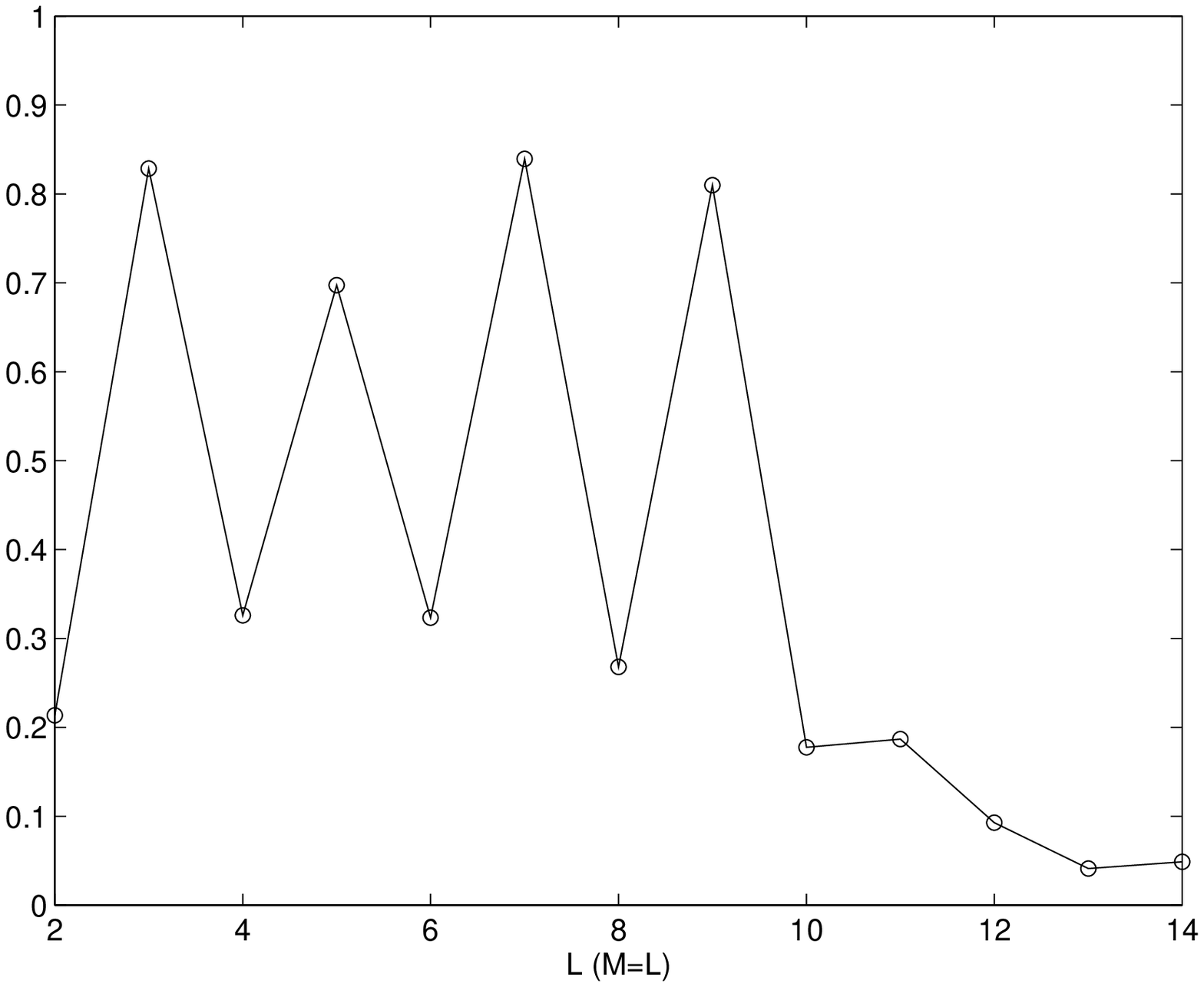 hoffset=-32 voffset=-84 hscale=45 
vscale=45}{2.97in}{2.4in}}
\newline
(a)\hspace{2.9in}(b)
\caption{$|\R_{11}|$ vs. $L$ for (a) $M=50$ and (b) $M=L$.  Note that,
as with the bound state problem,
setting $M=L$ produces unsatisfactory results.}
\label{RvsL}
\end{figure}

As with the bound state problem, we find that we must take
$M>L$ to achieve convergence.  To illustrate this further,
Figure \ref{RvsL} plots $|\R_{11}|$ for the same parameters, using
the method outlined in \S II.C, as a function of $L$ for $M=50$ and
for $M=L$.  From this, we can see clearly that the $M=L$ case is
unsatisfactory, as expected.  Although the speed of convergence varies
with the region of interest---it is significantly slower when the energy
is near a resonance---we have taken parameters which give accuracy to at
least a part in 100.

\section{Results and Discussion}

\subsection{Eigenenergies of the bound states}

\begin{figure}[htbp]
\centering
\PSbox{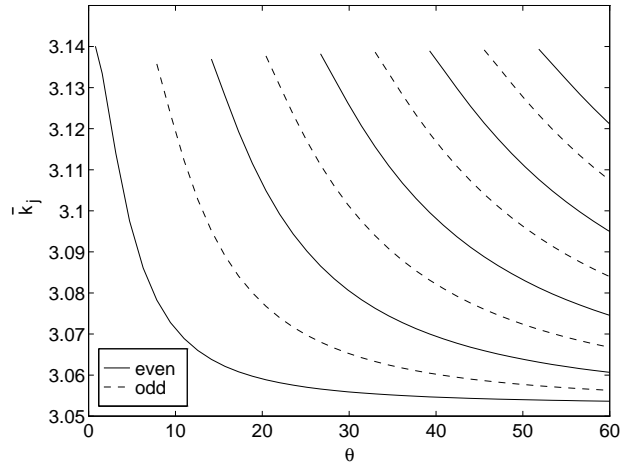 hscale=44 vscale=44 hoffset=-19 
voffset=-87}{3.1in}{2.5in}
\caption{Plot of $\bar{k_j}$ vs.\ $\theta$, for $\k = 0.8$.}
\label{eigenenergies}
\end{figure}

We first examine the dependence of the bound state energies on the
parameter $s_0$, the length of the curved region of the tube.  Figure
\ref{eigenenergies} plots the momenta $\bar{k_j}$ corresponding to bound state
eigenenergies against the angle of the circular arc $\theta$ $(= 2\k
s_0)$ for curvature $\kappa = 0.8$.  Eigenenergies for both symmetric
states and antisymmetric states are shown.

The curved region introduces a term that behaves like an attractive
finite square well potential with depth $\k^2/4$.  Increasing the
length of this curved region both lowers the energy of the ground
state and allows for more excited states.

\begin{figure}[htbp]
\centering
\parbox{2in}{\PSbox{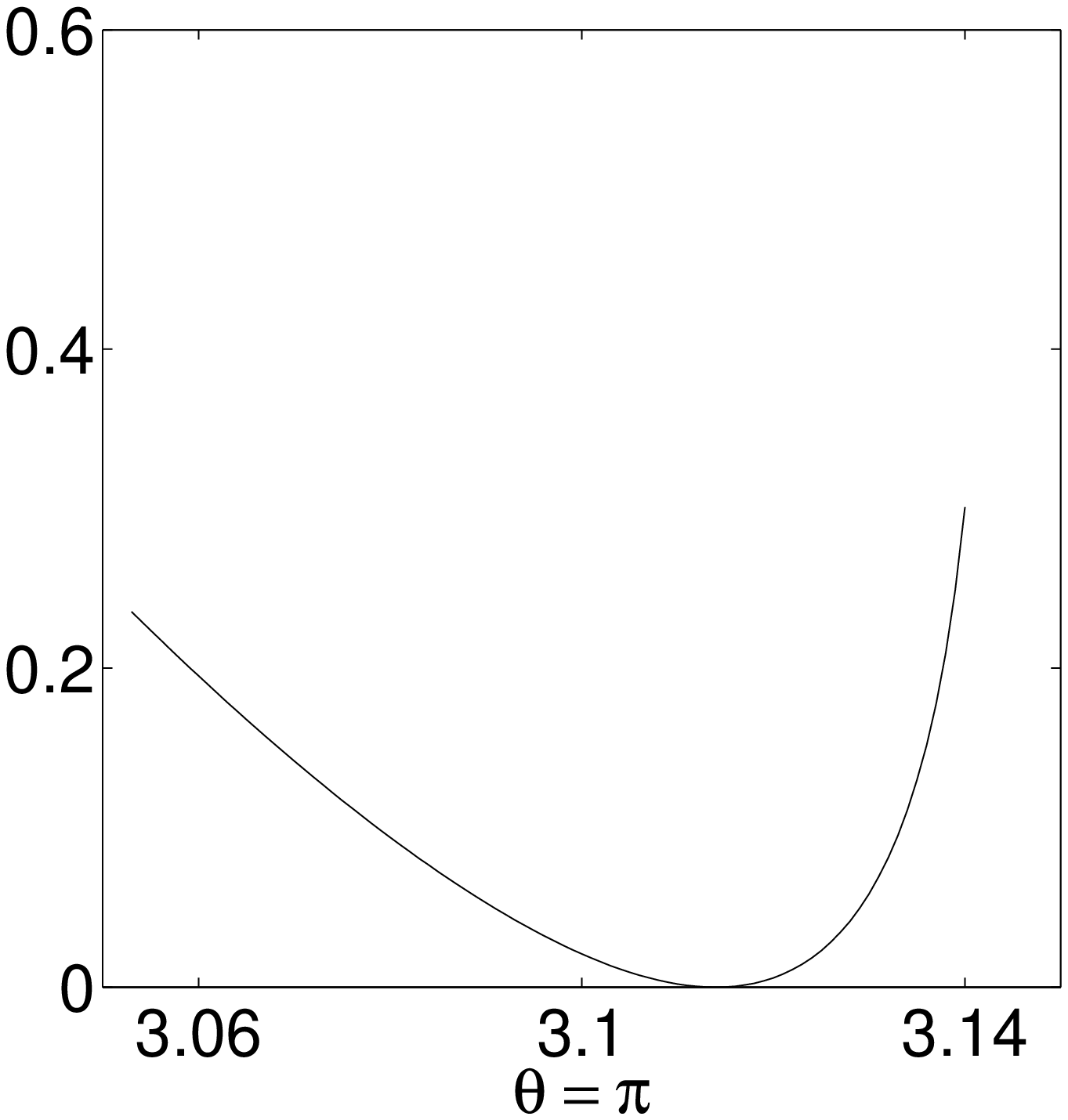 hoffset=-39 voffset=-58 hscale=36 vscale=36}
{1.95in}{2.1in}}
\parbox{2in}{\PSbox{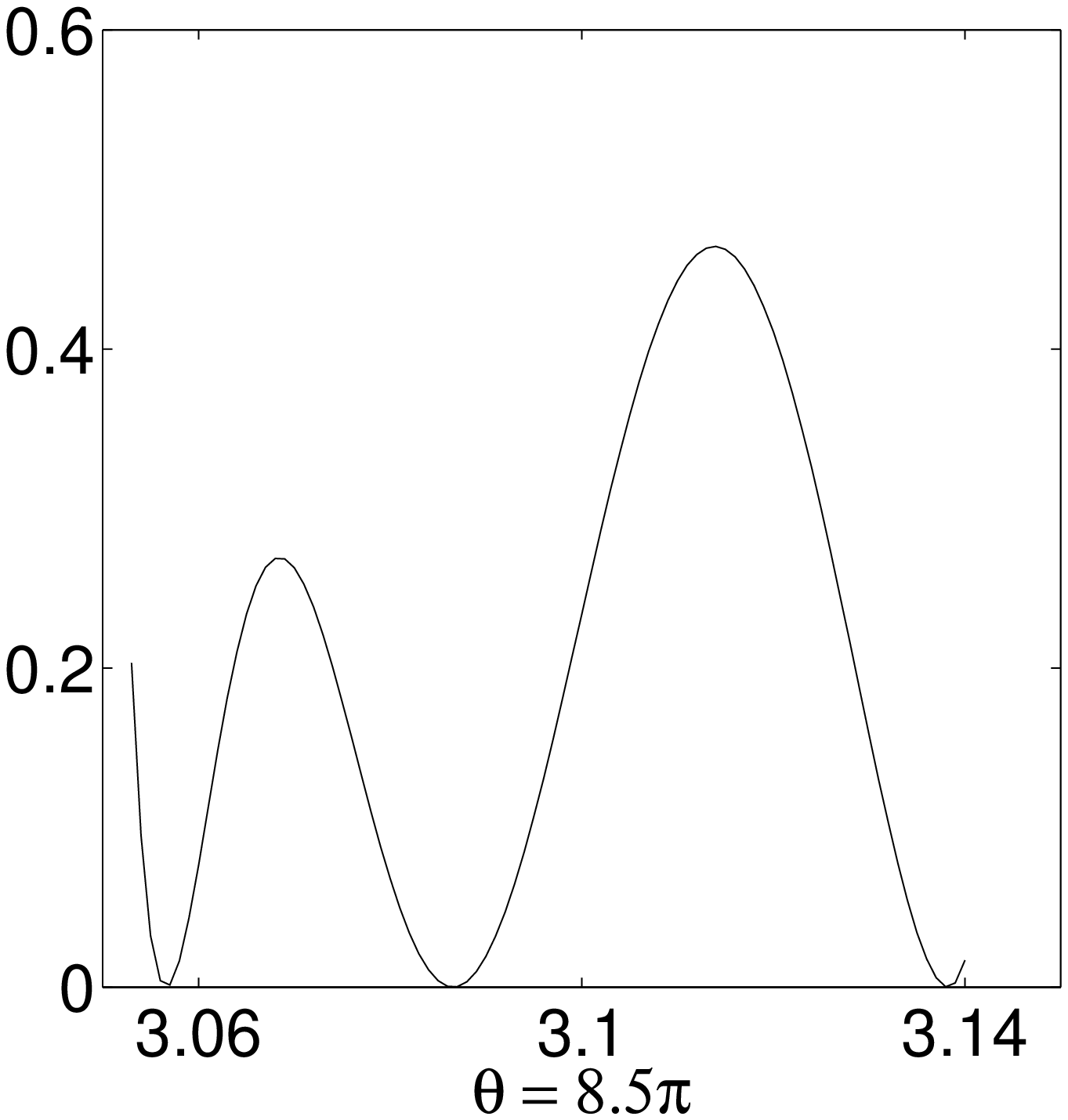 hoffset=-39 voffset=-58 hscale=36 vscale=36}
{1.95in}{2.1in}}
\parbox{2in}{\PSbox{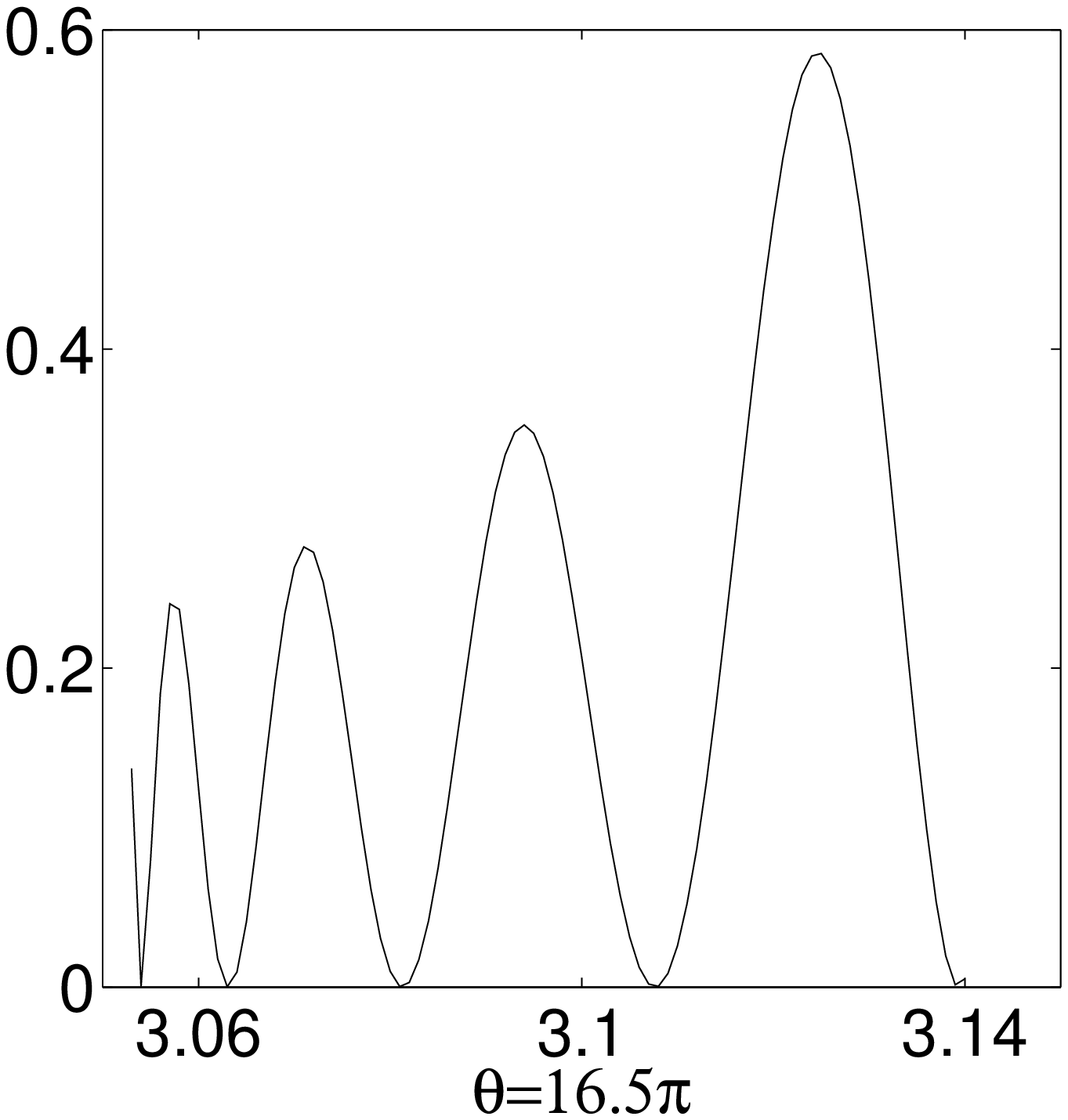 hoffset=-39 voffset=-58 hscale=36 vscale=36}
{1.95in}{2.1in}}
\caption{Plots of $\Delta (k)$
vs.\ $k$ for $\k = 0.8$ and different $\theta$.  Only symmetric states
are considered.  The minima correspond to $\bar{k_j}$ for odd values of
$j$.}
\label{multmin}
\end{figure}

Figure \ref{multmin} shows plots of the lowest eigenvalue of the
matrix $\Xi$ as defined in \S II.B against $k$, for curvature $\k
= 0.8$ and different values of the angle $\theta$.  The minima on
these graphs are candidate (symmetric) bound states provided $\bar k<\pi$, and
correspond to points in Figure \ref{eigenenergies} along a line
of fixed angle.  As the angle increases, the number of eigenstates
also increases.  

In the limit $\theta \rightarrow
\infty$, all the $\bar{k_j}$ approach the same value, which we will
denote by $\bar k_\infty$.
The value of $\bar k_\infty$ is that which corresponds
to $\nu = 0$ for any given curvature.  To see this, 
note that in the limit  $\theta \rightarrow \infty$, the external straight 
region becomes negligible (for bound states), and the internal curved region
becomes very long.   In this limit it becomes possible to accomodate
wavefunctions with any number  of nodes (in $s$) without introducing
significant energy.  The longitudinal  wavefunction, $\cos \kappa\nu s$,
(the antisymmetric case can be handled  analogously) can oscillate any
number of times over the range
$-s_0<s<s_0$  contributing only $\k^2\nu^2$ to the energy.  By choosing
$\nu$ very small  this contribution to the energy can be made negligible.  A
minute tuning of 
$\nu$ allows one to match the wavefunction to a falling exponential at 
$s=s_0$.  Thus as $s_0\rightarrow\infty$ there 
are an infinite number of bound 
states accumulating at the value of $k$ corresponding to $\nu=0$.  At any 
fixed very large $s_0$ there are also many other bound states at energies 
between $\bar k_\infty$ and $\pi$.  
The value of $\bar k_\infty$ is determined 
by solving the transcendental equation,
\begin{equation}
\left[Y_0\left(\frac{\bar k_\infty}{\k}\right)J_{0}\left(\frac{\bar k_\infty}
{\k}(1-\k)\right)
- J_{0}\left(\frac{\bar k_\infty}{\k}\right)
Y_{0}\left(\frac{\bar k_\infty}{\k}(1-\k)\right)\right] = 0.
\label{condition}
\end{equation}
This provides us with $\bar k_\infty(\k)$.  As $\k\rightarrow 1$, $Y_0$ 
becomes singular in eq.~(\ref{condition}).  A 
careful analysis of the behavior 
at small $1-\k$ shows that the equation reduces to $J_0(\bar 
k_\infty(\k=1))=0$, which gives $\bar k_\infty(1)=2.405\dots$.\cite{KL}

\begin{figure}[htbp]
\centering
\PSbox{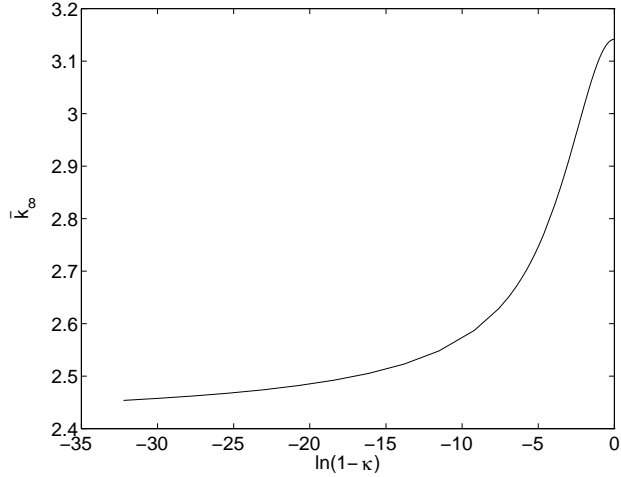 hscale=45 vscale=45 hoffset=-20 voffset=-80}{3.1in}{2.5in}
\caption{The value of $\bar k_\infty$ as a function
of $\ln (1-\k)$.}
\label{kK}
\end{figure}

Figure \ref{kK} plots the solution of eq.~(\ref{condition}),
$\bar k_\infty(\k)$ against $\ln (1-\k)$.  As $\k \rightarrow 0$ $(\ln (1-\k)
\rightarrow 0)$, the tube is straight and so $\nu = 0$ corresponds to $k
= \pi$, as expected.  And as $\k\rightarrow 1$, $\bar 
k_\infty\rightarrow 2.405\dots$ as expected.
It should be noted that the amount of binding increases dramatically as 
$\k\rightarrow 1$, a fact 
which necessitated our choice of $\ln (1-\k)$ as the 
independent variable in Fig.~\ref{kK}.

Next we turn our attention to determining the angle at which the first 
(antisymmetric) bound excited state appears.
We label the angles at which new bound states appear as $\bar{\theta_1}$, 
$\bar{\theta_2}$, {\em etc.}\ where $\bar{\theta_1}$ corresponds to the
ground state and $\bar{\theta_j}$ $(j>1)$ to the $(j-1)^{th}$ excited state.  
For any $\k$, $\bar{\theta_1}$ is zero, since there exists a bound
state for any tube with a curved region, no matter how short.  

\begin{figure}[htbp]
\centering
\PSbox{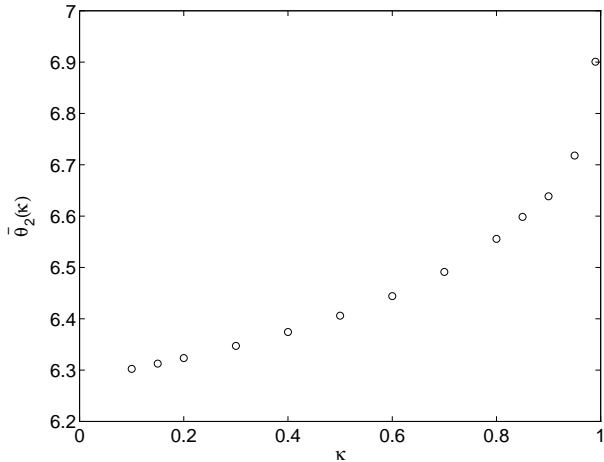 hscale=44 vscale=44 hoffset=-16 voffset=-80}{3.1in}{2.44in}
\caption{The angle $\theta$ at which the first antisymmetric bound
state appears as a function of curvature $\k$.  As $\k \rightarrow 0$,
$\theta \rightarrow 2\pi$.}
\label{antisym}
\end{figure}

Figure \ref{antisym} shows a plot of $\bar{\theta_2} (\k)$.
As can be seen from the graph, $\bar{\theta_2}$ approaches $2\pi$ as
$\k$ approaches zero.  We can see that this must be the case by
considering the one-channel approximation to the interior 
wavefunction, namely

\begin{equation}
\psi = \sin \pi y \sin \nu_1 \k s.
\label{wfxn}
\end{equation}
It is sufficient to consider this approximation because in the limit
that $\k\rightarrow 0$, the tube is very close to straight and thus
the contributions from higher channels are negligible.  In this limit,
the problem reduces to that of a square well potential.  The
Schr\"odinger equation then requires

\begin{equation}
\nu_1^2\k^2 - \frac{\k^2}{4} = k^2 - \pi^2.
\label{BScond}
\end{equation}

At the angle where the bound state first appears, $k=\pi$ so
eq.~(\ref{BScond}) implies $\nu_1 = \frac{1}{2}$.  For a bound state
to exist, the interior wavefunction ($\sin \nu_1\k s$) must be able to
match to a falling exponential at the boundary $s=s_0$.  The smallest
value of $s_0$ at which this can occur is that which satisfies the
condition $\nu_1\k s_0 = \pi/2$.  Recalling that $s_0 =
\theta/2\k$, we then have

\begin{equation}
\frac{\k}{2} \left( \frac{\theta}{2\k} \right) 
= \frac{\pi}{2} \Longrightarrow \theta = 2\pi.
\end{equation}

We note that as $\k$ is increased, the value of $\bar{\theta_2}$ also
increases.  This seems somewhat counter-intuitive, as we expect from the
Schr\"odinger equation that higher curvature implies greater binding, so
that the length of the curved region necessary to bind an excited state
should {\em decrease} with large $\k$.  This is indeed the case.  However, we
must recall that $\theta$ depends on $\k$ as well as $s_0$.  For a fixed
value of $\theta$, $s_0$ is much larger for small $\k$ than for large
$\k$.  Thus, the attractive potential acts over a much longer distance
for small $\k$, and this effect overshadows the difference in binding energy
due to differences in $\k$ alone.


\subsection{Scattering results}

Previous workers have noted many features of the scattering problem, and we 
do not have any qualitatively new phenomena to add to their 
treatments.\cite{Lent,SM}  The primary features of interest to us are the 
rapid rise of $\vert \T_{11}\vert$ from threshold, the steady increase of the 
phase of $\T_{mm}$ in regions of 
energy away from thresholds, and finally, the 
dramatic fluctuations in $\T$ and $\R$ at energies just below thresholds.
Our interest lies in understanding better the 
physical origins of these effects.  
These features are due to the 
presence of a bound state just below threshold, a steadily growing phase
over a wide range of $k$, and weakly coupled
quasibound states just  below each channel threshold, respectively.  The
last of these provides an  elegant example of a resonance manifesting itself
as total destructive  interference --- a phenomenon familiar to particle
physicists in the form of  the sudden drop in the $\pi\pi\rightarrow\pi\pi$
$S$-wave cross section just  below $\bar K K$ threshold, due to the presence
of the
$a_0$  resonance.\cite{PH}

\begin{figure}
\centering
\PSbox{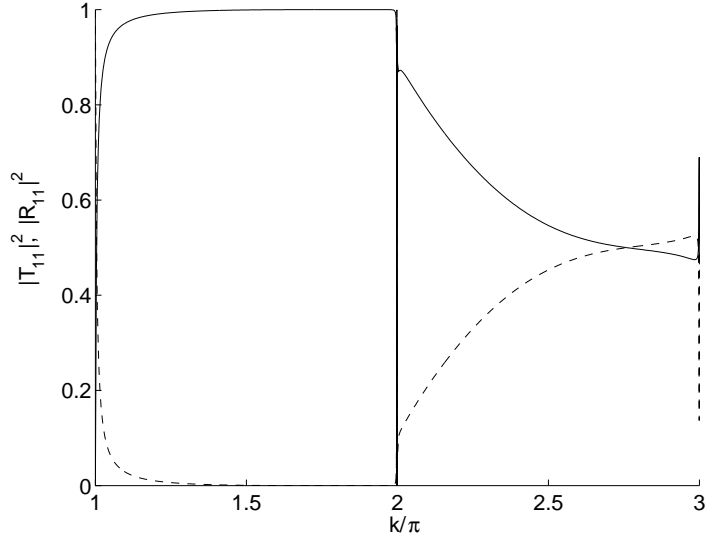 hoffset=-35 voffset=-100 hscale=51 vscale=51}{3.1in}{2.7in}
\caption{$|\T_{11}|^2$ (solid) and $|\R_{11}|^2$ 
(dashed) vs. $k/\pi$.  Here $\k=0.9$, 
$\theta = \pi/2$.}
\label{TR}
\end{figure}

We begin by examining $|\T_{11}|$ and $|\R_{11}|$ as functions of $k$.  
Figure \ref{TR} plots $|\T_{11}|^2$ and $|\R_{11}|^2$, the measures of 
transmitted and reflected flux, for $1 \leq N \leq 3$ at fixed
$\k = 0.9$ and $\theta = \pi/2$.

Note that in 
the region where $N=1$ ($\pi \leq k < 2\pi$), 
$|\T_{11}|^2 + |\R_{11}|^2 = 1$.  
This is required by unitarity and serves as a 
check on our numerical calculations.
We have checked that our $\CS$-matrix remains unitary in the 
$N\times N$ case as well.  As 
$k$ crosses the second threshold at $2\pi$ there 
is significant inelasticity into the newly open channel evidenced by the fact 
that $|\T_{11}|^2 + |\R_{11}|^2 < 1$.  

We also observe that at energies just below $N^2\pi^2$,
the transmission amplitude in the highest open channel ($\T_{N-1\,N-1}$)
drops sharply almost to zero.
We shall show that this and other rapid variations in $\T$ and $\R$ are
manifestations of resonances related
to the presence of quasibound states just below 
each new channel threshold.  They are the energies at which
we would find a bound state if all open channels were artificially
closed.  

\begin{figure}
\centering
\parbox{3in}{\PSbox{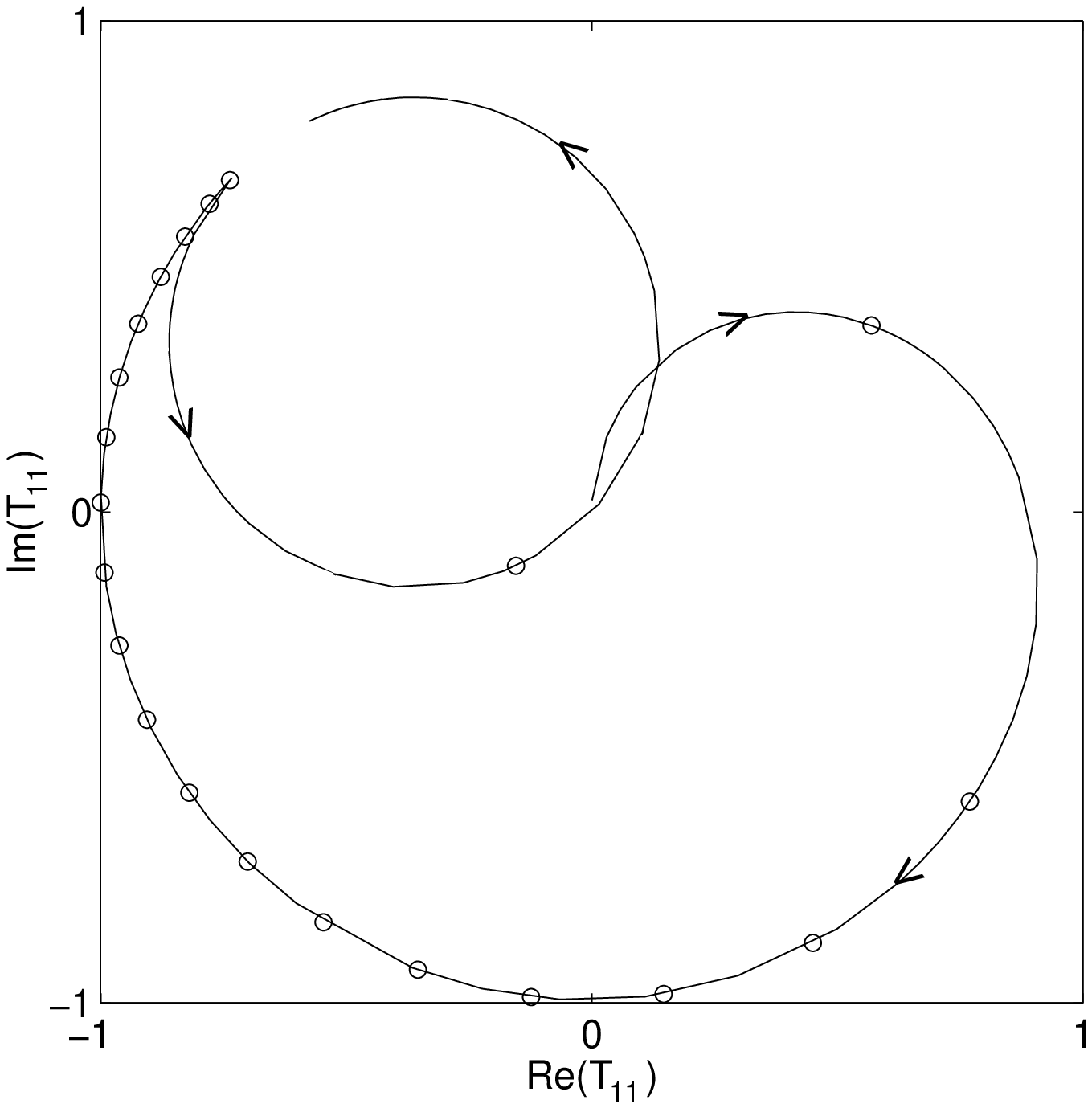 hoffset=-55 voffset=-90
hscale=53 vscale=53}{2.9in}{3.0in}}
\parbox{3in}{\PSbox{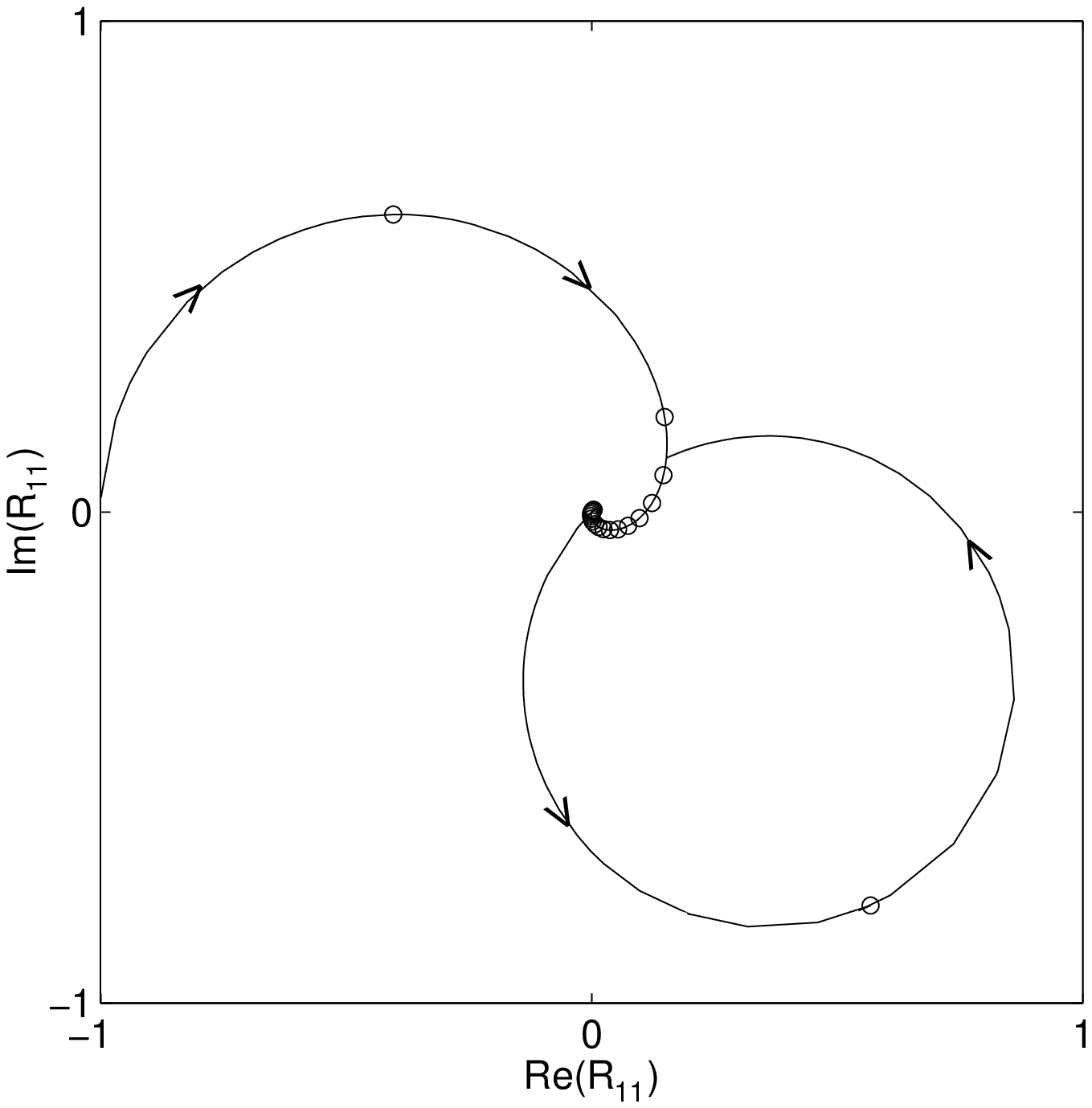 hoffset=-55 voffset=-90
hscale=53 vscale=53}{2.9in}{3.0in}}
\newline
(a) \hspace{2.8in} (b) \\
\caption{Argand diagrams for (a) $\T_{11}$ and (b) $\R_{11}$
for the parameters $\k=0.9$, $\theta = \pi/2$, $\pi < k < 2\pi$.
Arrows indicate direction of increasing $k$; circles mark equal
intervals in $k$.}
\label{argand}
\end{figure}

It is difficult to figure out what is happening by considering $|\T|$ and 
$|\R|$ alone.  We will learn much more by looking at the Argand diagrams for 
$\T$ and $\R$.  The general 
characteristics will be illustrated by considering 
plots of $\T_{11}$ and $\R_{11}$ in the complex plane for $\pi < k < 2\pi$
as shown in figure \ref{argand}.

$\RR$ and $\TT$ are not independent when only one 
channel is open.  Unitarity ($\CS^\dagger\CS=1$) requires both 
$|\TT|^2+|\RR|^2=1$ and ${\rm Re}\,\TT^*\RR=0$.  We 
can parameterize both by the 
modulus ($M$) and phase ($\phi$) of $\TT$,
\begin{eqnarray}
\TT&=&Me^{i\phi}\nonumber\\
\RR&=&i\sqrt{1-M^2}e^{i\phi}
\label{unit}
\end{eqnarray}
So it suffices to discuss the behavior of $\T_{11}$.  According to
Fig.~(\ref{argand}), $\T_{11}$ vanishes at $k=\pi$.  As $k$ increases,
$\TT$ rapidly executes a clockwise circle of radius $1/2$ in the
complex plane until $|\TT|\approx 1$ at a $k$ only slightly larger
than $\pi$.  Then $\TT$ moves clockwise on the unit circle 
until suddenly, at a $k$ just below $2\pi$, $\TT$ suddenly
executes an almost complete counterclockwise circle of radius $1/2$.
$\TT$ briefly resumes its steady phase growth, until the second
channel opens at $k=2\pi$ and $\TT$ moves off the unit circle.

First consider the behavior near $k=\pi$.  We know on general grounds 
that $\TT=0$ and $\RR=-1$ at threshold.  Furthermore, we 
know that there is a symmetric 
bound state at an energy $k^2 = \bar k_1^2=\pi^2-\varepsilon$.  
This must appear as a pole in $\T + \R$ at this energy.  Since $\CS$ must 
be unitary for any real $k^2$, $\TT+\RR$ must be of the form,
\begin{equation}
\TT+\RR={k_1+i\varepsilon\over{k_1 -
i\varepsilon}}+f_\sigma(k)
\end{equation}
where $k_1=\sqrt{k^2-\pi^2}$, and 
$f_\sigma(k)$ is a smoothly varying function of $k$ near 
$k=\pi$.  There is no nearby antisymmetric bound state, so $\TT-\RR$ is a 
smooth function of $k$,
\begin{equation}
\TT-\RR=f_\alpha(k)
\end{equation}
near $k=\pi$.  Comparing with the 
threshold behavior, $\TT=0$ and $\RR=-1$, we 
see that $f_\sigma(\pi)=0$ and $f_\alpha(\pi)=1$.  
Combining all these results 
we obtain,
\begin{eqnarray}
\TT &=& e^{i\phi}\cos\phi\nonumber\\
\RR &=& ie^{i\phi}\sin\phi\quad{\rm where}\nonumber\\
\tan\phi &\equiv& {\varepsilon\over k_1}\nonumber\\
\label{thresh}
\end{eqnarray}
At threshold, $\phi=\pi/2$, and as $k$ increases, 
$\phi$ drops rapidly to zero 
over an interval scaled by $\varepsilon$.  
Eqs.~(\ref{thresh}) parameterize a 
semicircle of radius $1/2$ centered at $\TT=1/2$, which $\TT$ executes as 
$k_1$ grows from zero to a value greater than $\varepsilon$.
The speed at which $\TT$ moves measures $\varepsilon$, by the formula
\begin{equation}
\varepsilon=ik_1\left({1\over\TT}-1\right)
\end{equation}

For $k_1\gg\varepsilon$ the conditions of the small $\k$
approximation 
discussed in \S II are satisfied, 
provided $\k$ is small.  Thus we expect 
to find $\RR\approx 0$ and $\TT=\TT^0\times\exp(-ik_1\theta/2)$.  
We verify this by comparing the rate of change of the argument of
$\TT$ with the small $\k$  prediction, $d\arg\TT/dk_1 =- \theta/2$.  
So in the
range where the approximation is valid, we expect that 
the graph of the phase angle
$\phi$ of $\T$ will be linear with respect to $k_1$, with a slope
equal to half the angle subtended by the curved region of the tube.
Computing $\T_{11}$ for $4.0<k_1<5.0$ with 
$\kappa = 0.001$ and $\theta=\pi/2$,
a range and curvature for which the small $\k$ approximation 
is valid, we have found
$d\arg\TT/dk_1 \approx -\pi/4 = -\theta/2$, which confirms that this 
relation is indeed satisfied.

Finally we turn to the dramatic behavior of $\T$ and $\R$ near channel 
thresholds.  It is clear from Fig.~(\ref{argand}) that $\TT$ resonates ({\it 
i.e.\/} it rapidly executes a counterclockwise circle in the complex 
plane\cite{GW}) at the energy where $|\TT|$ 
drops precipitously to zero.  This 
behavior is indicative of a pole in the complex $k^2$ plane just below the 
real axis.  The pole appears in $\TT+\RR$ because the quasibound state is 
symmetric in $s$.  $\TT-\RR$ is smooth over this region in $k$.  If we denote 
the pole location as $q_1-i\gamma_1$, where $q_1\lessapprox 2\pi$ and 
$\gamma_1$ is small and positive (by causality). 
Then in the neighborhood of the pole,
\begin{equation}
\TT+\RR={k_1-q_1-i\gamma_1\over k_1-q_1+i\gamma_1} + \ldots,
\end{equation}
where $\ldots$ denotes other, smooth terms in $\TT+\RR$.  Combining this with 
a smooth form for $\TT-\RR$ we see that $\TT$ executes a counterclockwise 
circle of radius $1/2$ in the complex $k$-plane as $k$ passes over the pole.  
Just before the vicinity of the pole, $\TT$ lay nearly on the unitarity 
circle, $|\TT|\approx 1$.  The only way that it could rapidly execute a 
counterclockwise circular path consistent with unitarity 
is if the resonant circle is {\it tangent\/} 
to the unitarity circle.  [Otherwise $\TT$ 
would have to pass outside the unit 
circle at some point on its circular path.]  
Since the radius of the resonance 
circle is $1/2$, the (tangent) resonant circle {\it must pass through the 
origin\/}.  Thus $|\TT|$ must vanish in the near vicinity of the quasibound 
state as observed by Ref.~\cite{SM}, and shown in Fig.~(\ref{argand}).

\begin{figure}
\centering
\parbox{3in}{\PSbox{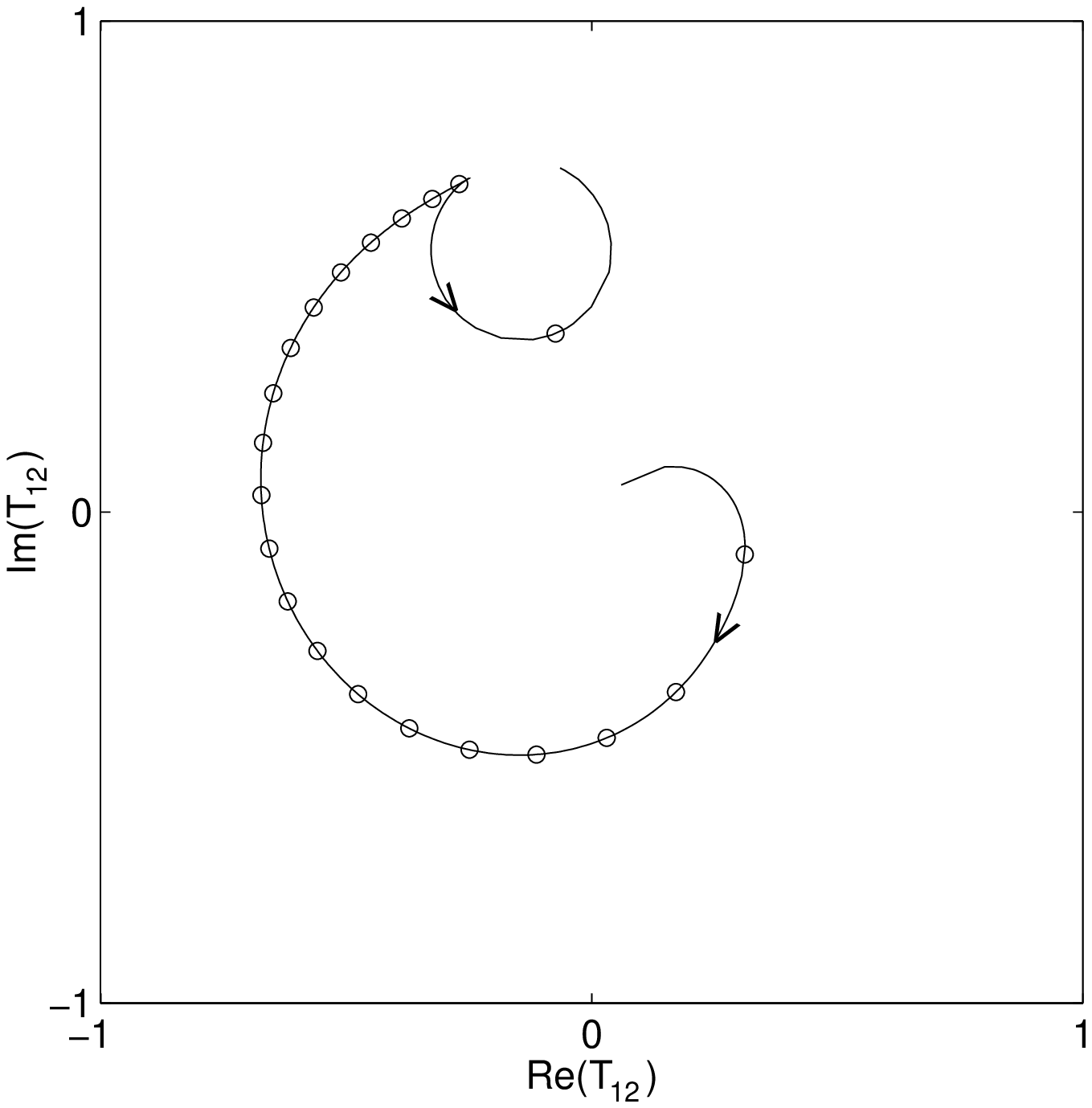 hoffset=-60 voffset=-90
hscale=53 vscale=53}{2.9in}{3.1in}}
\parbox{3in}{\PSbox{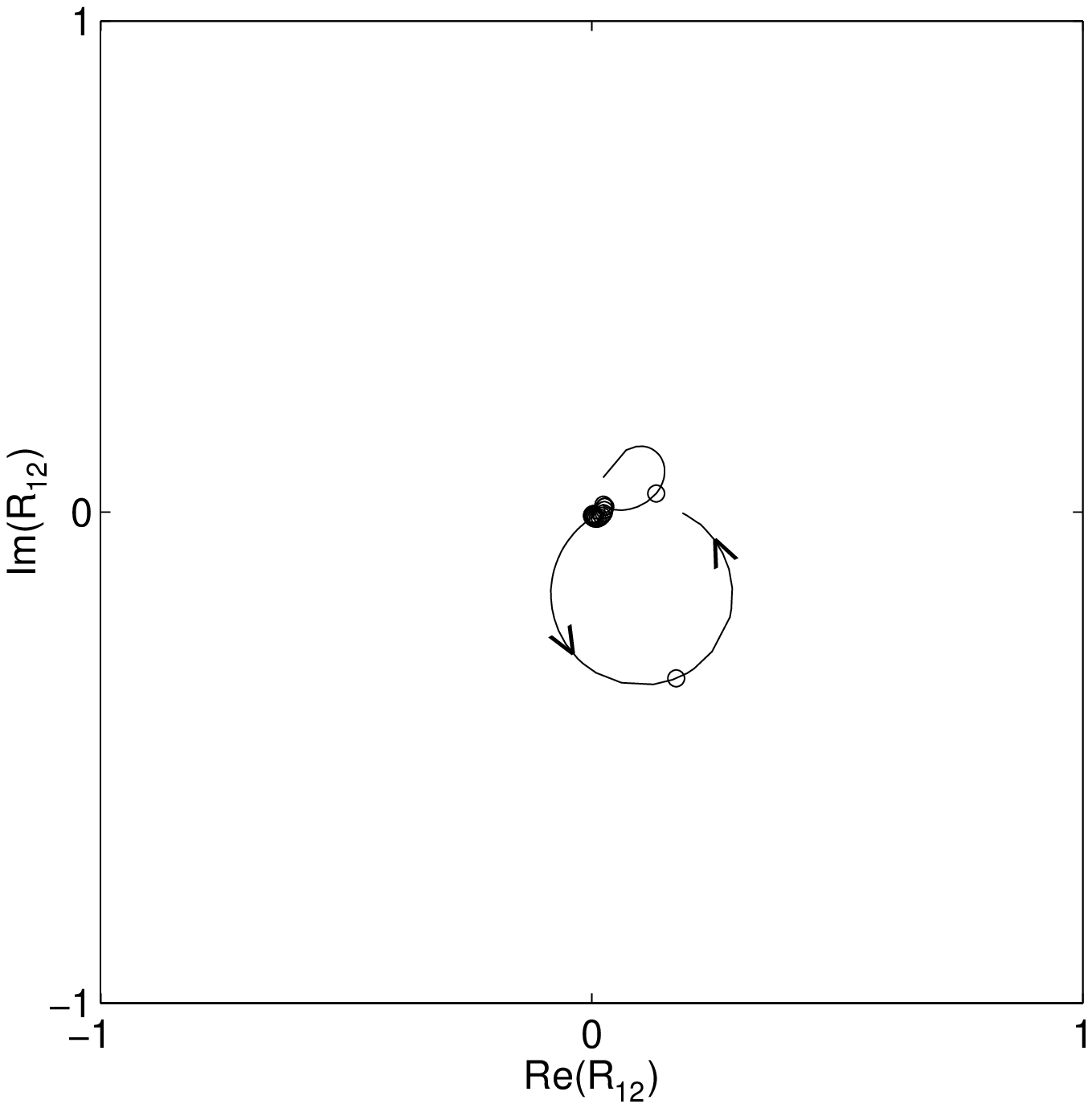 hoffset=-60 voffset=-90
hscale=53 vscale=53}{2.9in}{3.1in}}\\
\parbox{3in}{\PSbox{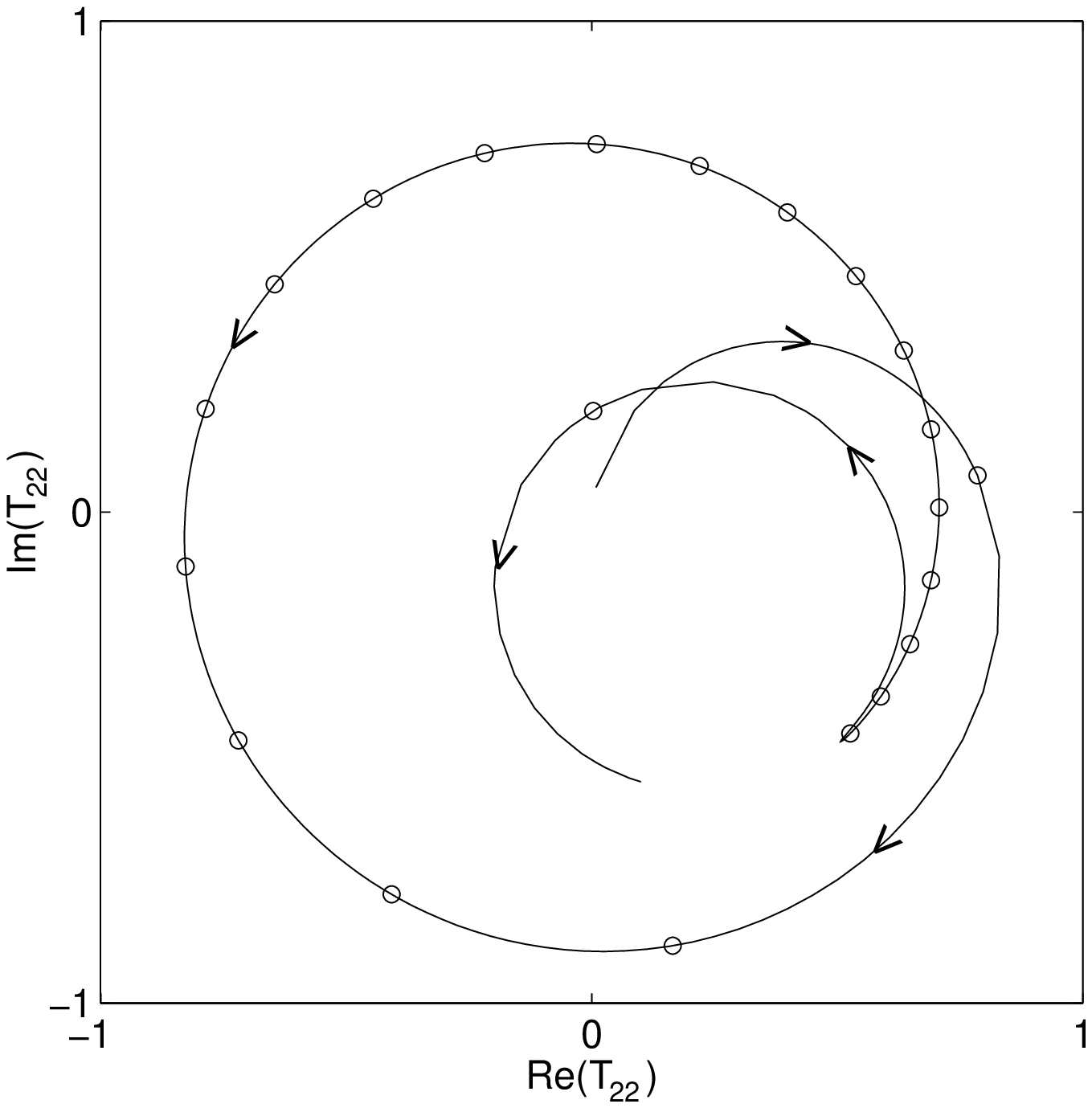 hoffset=-60 voffset=-95
hscale=53 vscale=53}{2.9in}{2.9in}}
\parbox{3in}{\PSbox{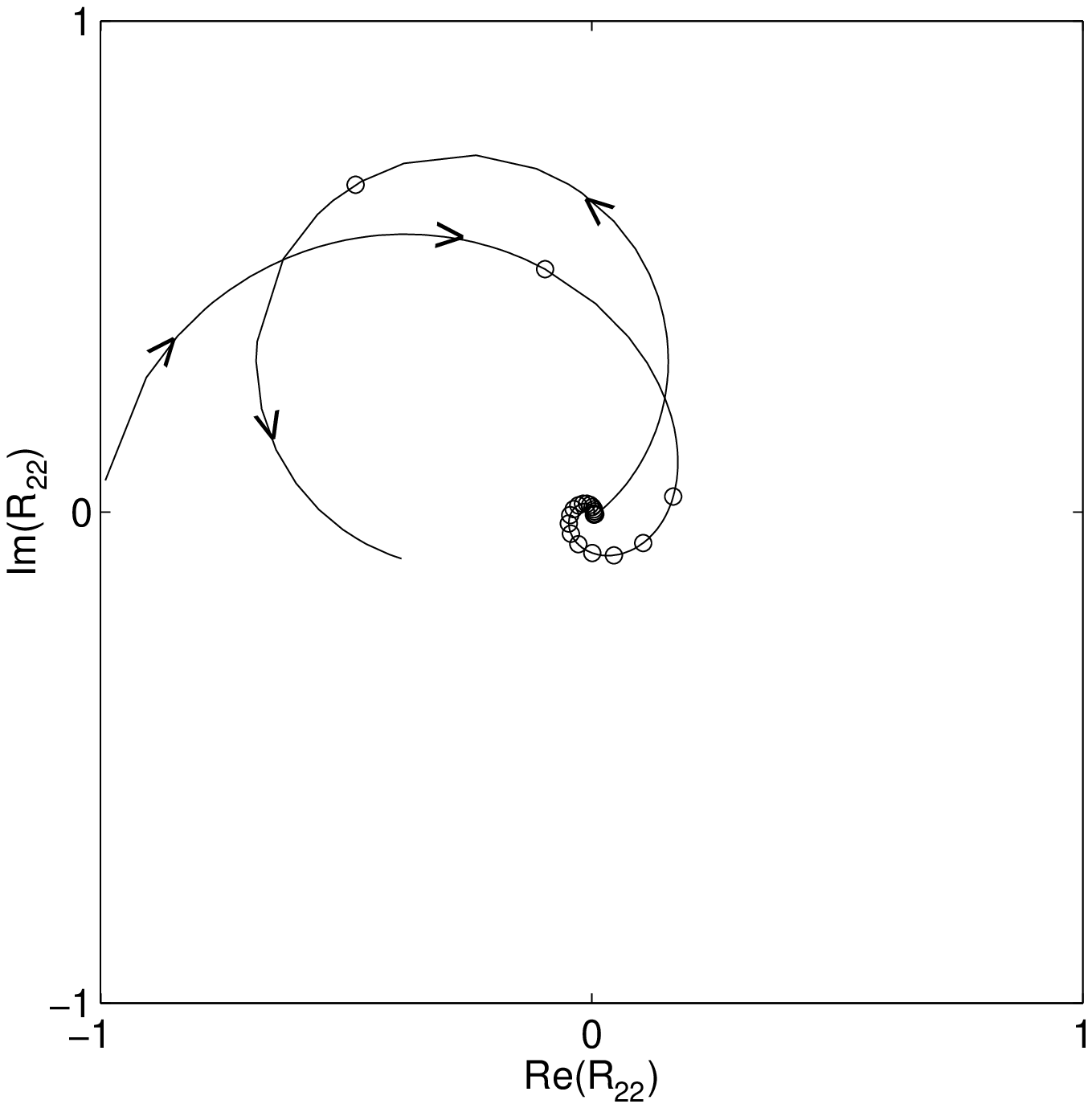 hoffset=-60 voffset=-95
hscale=53 vscale=53}{2.9in}{2.9in}}\\
\caption{Argand diagrams for $\T_{12} = \T_{21}$, $\T_{22}$, 
$\R_{12} = \R_{21}$, and $\R_{22}$ for the range $2\pi < k < 3\pi$.
The same basic structure is observed as for the coefficients in the
$N=1$ case.}
\label{argand2}
\end{figure}

The basic structure of $\T_{11}$ and $\R_{11}$ can be seen in the
scattering coefficients for higher channels.  As we observe in figure
\ref{argand2} for $N=2$, $\T_{12}$ and $\T_{22}$ follow the same pattern.
For $k_2 = \sqrt{k^2 - 4\pi^2}$ just above zero, $\T$ executes rapid
clockwise motion as a result of the pole located near $k = \pi$.  This
is followed by smooth phase growth until $k$ reaches the quasi-bound
state just under the $N=3$ threshold, at which time it executes the
rapid counterclockwise circle in the complex plane characteristic of
an elastic resonance. 

Our conclusion from this exercise is that the behavior observed in
Fig.~\ref{argand} is characteristic of scattering in bent tubes.  The
bound states, quasibound states and regions of smooth phase growth are
robust properties of these systems.  They do not depend specially on
our choice of a circular geometry which facilitated our numerical
calculations.  Perhaps the most interesting observation is the
appearance of violent fluctuations in transmission and reflection
properties of bent tubes in the vicinity of channel thresholds.  The
fluctuations occur over very narrow intervals in energy (the
resonances are narrow), so low resolution experiments would quite
likely fail to detect them.  On the other hand, careful experiments
using, for example, microwave radiation in waveguides, should see
these rapid fluctuations in response associated with the opening of
channel thresholds in which quasibound states occur.

\end{document}